\documentclass[journal,draftclsnofoot,onecolumn,12pt]{IEEEtran}

\usepackage{amsthm,amssymb,graphicx,multirow,amsmath,color,amsfonts}%,ulem}
\usepackage[update,prepend]{epstopdf}
\usepackage[noadjust]{cite}
\usepackage[latin1]{inputenc}
\usepackage{tikz}
\usetikzlibrary{arrows,calc}		% Optional ticks libraries
\usepackage{bbm} % for \mathbbm{1}
\usepackage{pdfpages}
\usepackage{tabulary}
\usepackage{multirow}
\usepackage{comment}
\usepackage{dsfont}
\usepackage{pgf,tikz}
\usepackage{mathrsfs}
\usetikzlibrary{arrows}
\usepackage{subcaption}
\usepackage{bigints}
\usepackage{mathtools}
\usepackage{dsfont}
\usepackage{color}

\usepackage{algorithmic}

\newcommand\numberthis{\addtocounter{equation}{1}\tag{\theequation}}
\usepackage[]{algorithm2e}
\SetKwInput{KwInput}{Input}
\SetKwInput{KwOutput}{Output}
\DeclareMathOperator*{\argmax}{arg\,max}

\newcommand{\cdf}{\mathtt{CDF}}

\def\nbc{{\mathbf{c}}}

\def\nbf{{\mathbf{f}}}
\def\nbg{{\mathbf{g}}}
\def\nbh{{\mathbf{h}}}

\def\nbo{{\mathbf{o}}}
\def\nbp{{\mathbf{p}}}

\def\nbr{{\mathbf{r}}}

\def\nbu{{\mathbf{u}}}

\def\nbw{{\mathbf{w}}}
\def\nbx{{\mathbf{x}}}
\def\nby{{\mathbf{y}}}

\def\nb0{{\mathbf{0}}}
\def\nb1{{\mathbf{1}}}

\def\nbI{{\mathbf{I}}}

\def\ncalA{{\mathcal{A}}}

\def\ncalE{{\mathcal{E}}}

\def\ncalH{{\mathcal{H}}}
\def\ncalI{{\mathcal{I}}}

\def\ncalL{{\mathcal{L}}}
\def\ncalM{{\mathcal{M}}}

\def\ncalP{{\mathcal{P}}}

\def\ncalS{{\mathcal{S}}}

\def\ncalV{{\mathcal{V}}}
\def\ncalW{{\mathcal{W}}}
\def\ncalX{{\mathcal{X}}}
\def\ncalY{{\mathcal{Y}}}

\def\nbbC{{\mathbb{C}}}

\def\nbbE{{\mathbb{E}}}

\def\nbbP{{\mathbb{P}}}

\newtheorem{lemma}{Lemma}

\newtheorem{remark}{Remark}

\def\argmax{\operatorname{arg~max}}
\def\R{\mathbb{R}}

\def\sinr{\mathtt{SINR}}			% Signal to interference plus noise ratio
\def\snr{\mathtt{SNR}}

\allowdisplaybreaks % Allows breaking of eqnarray over multiple pages (avoids unnecessary blanks in the document before eqnarray)

\begin{document}
\title{Pilot Assignment Schemes for Cell-Free Massive MIMO Systems}
\author{
Priyabrata Parida and Harpreet S. Dhillon \vspace{-0.75cm}
\thanks{Authors are with Wireless@VT, Dept. of Electrical and Computer Engineering, Virginia Tech, Blacksburg, VA (Email: \{pparida, hdhillon\}@vt.edu). The support of the US NSF (Grant ECCS-1731711) is gratefully acknowledged. This paper was presented in part at the IEEE Globecom, 2019~\cite{Parida2019gc}.} % remove the date for conference drafts
}

\maketitle

\vspace{-1cm}

\begin{abstract}
In this work, we propose three pilot assignment schemes to reduce the effect of pilot contamination in cell-free massive multiple-input-multiple-output (MIMO) systems. Our first algorithm, which is based on the idea of random sequential adsorption (RSA) process from the statistical physics literature, can be implemented in a distributed and scalable manner while ensuring a minimum distance among the co-pilot users. Further, leveraging the rich literature of the RSA process, we present an approximate analytical approach to accurately determine the density of the co-pilot users as well as the pilot assignment probability for the typical user in this network. We also develop two optimization-based centralized pilot allocation schemes with the primary goal of benchmarking the RSA-based scheme. The first centralized scheme is based only on the user locations (just like the RSA-based scheme) and partitions the users into sets of co-pilot users such that the minimum distance between two users in a partition is maximized. The second centralized scheme takes both user and remote radio head (RRH) locations into account and provides a near-optimal solution in terms of sum-user spectral efficiency (SE). The general idea is to first cluster the users with similar propagation conditions with respect to the RRHs using spectral graph theory and then ensure that the users in each cluster are assigned different pilots using the branch and price (BnP) algorithm. Our simulation results demonstrate that despite admitting distributed implementation, the RSA-based scheme has a competitive performance with respect to the first centralized scheme in all regimes as well as to the near-optimal second scheme when the density of RRHs is high.
\end{abstract}
\vspace{-0.5cm}
\vspace{-0.1cm}
\section{Introduction}
{The concept of distributed implementation of MIMO technique has been actively explored for more than a decade. 
%However, it was not up until recently, when its practical implementation appeared more viable due to densification of wireless networks.
Recent push towards network densification has also made its practical implementation viable.}
In a distributed MIMO network, a number of geographically separated RRHs simultaneously serve users in the network and are connected to a central base band unit (BBU), which performs the physical layer tasks of symbol detection and decoding along with complex upper layer tasks such as resource scheduling.
In a dense network, the access points (or RRHs) can be leveraged for distributed implementation resulting in higher SE through macro-diversity and elimination of frequent handovers as users move from one access point to another.
Apart from network densification, massive MIMO (mMIMO) is also one of the major enablers of the 5G wireless networks.
Hence, it is natural to study the performance of a wireless network where a large number of RRHs with multiple antennas simultaneously serve multiple users.
Not surprisingly, such studies have been conducted over the years under different pseudonyms such as network MIMO, {coordinated multi-point}~\cite{Irmer2011}, cloud radio access network (RAN)~\cite{Checko2015}, fog-mMIMO~\cite{Bursalioglu2019}, and recently the cell-free mMIMO~\cite{Nayebi2017, Ngo2017}.
Successful implementation of any MIMO technique requires accurate channel state information (CSI) at the transmitter (or receiver). 
Similar to cellular mMIMO networks, the CSI acquisition in cell-free mMIMO systems needs to be done through uplink pilot transmission due to its scalability.
{However, under the assumptions of independent Rayleigh fading and sub-optimal linear precoders, {\em pilot contamination} becomes the only capacity limiting factor of both cellular and cell-free mMIMO networks~\cite{Marz2010, Ngo2017, Emil2018}.}
Hence, judicious pilot assignment is essential to reduce the effect of pilot contamination, which is the main focus of this paper.

\subsection{Related works}
In general, the optimal pilot assignment problem for a cell-free massive MIMO system is non-deterministic polynomial-time (NP)-hard in nature. 
Hence, the computational resources required to obtain the optimal solution scale exponentially with the number of users.
Therefore, almost all the works in the literature focus on providing heuristics-based algorithms to get an efficient solution.
These algorithms can be broadly categorized into centralized and distributed schemes.
In~\cite{Ngo2017}, a distributed random pilot allocation and a centralized greedy pilot allocation schemes are presented for a cell-free mMIMO network.
In~\cite{Bursalioglu2019} and \cite{Bursalioglu2016}, a distributed random access type pilot assignment scheme is proposed, where a user is not served if its channel state information (CSI) cannot be estimated reliably.
A centralized structured pilot allocation scheme with an iterative application of the K-means clustering algorithm is presented in~\cite{Attarifar12018}. 
A natural way to address the resource allocation problem is through graph theoretic framework.
This idea has been explored in~\cite{Zhang2017, hmida2020graph, Zeng2021}.
In~\cite{Zhang2017}, a centralized pilot sequence design scheme is proposed where the users in the neighborhood of an RRH use orthogonal pilot sequences.
The problem is posed as a vertex coloring problem and solved using the greedy DASTUR algorithm.
Along the similar lines, the authors of~\cite{hmida2020graph} construct the conflict graph by having an edge between users that are dominant interferers to each other.
The graph coloring problem is solved using a greedy algorithm.
In~\cite{Zeng2021}, the pilot assignment problem is mapped to the Max $K$-cut problem, which is solved using a heuristic algorithm.
In~\cite{Sabbagh2017}, a dynamic pilot allocation approach is presented where two users can be assigned the same pilot sequence if the signal to interference and noise ratios ($\sinr$s) of both the users are above a certain threshold.
While the aforementioned works primarily focus on reducing the interference due to pilot contamination, authors in~\cite{Liu2020} and \cite{buzzi2020pilot} solve pilot allocation optimization problems to maximize certain utility metrics.
Due to the NP-hard nature of the problem, authors in~\cite{Liu2020} use the Tabu search to solve the pilot allocation problem with the objective of maximizing sum-user SE.
Further, in~\cite{buzzi2020pilot}, system throughput maximization, and minimum user throughput maximization problems are solved using an iterative scheme based on the Hungarian algorithm.
Most of these works rely on the common underlying principle that the same pilot can be assigned to the users that have sufficient geographical separation. This principle also motivates the main pilot assignment scheme proposed in this work along with the additional objective that it should be distributed as well as scalable in nature while providing competitive performance in terms of average user SE. Our contributions are summarized next.
\subsection{Contributions}
{\em 1. Random sequential adsorption (RSA)-based pilot assignment scheme:} First, we propose a random pilot assignment algorithm with a minimum distance constraint among the co-pilot users to reduce the effect of pilot contamination.
The algorithm is inspired by random sequential adsorption (RSA) process, {which has been traditionally used across different scientific disciplines such as condensed matter physics, surface chemistry, and cellular biology, to name a few, to study the adsorption of large-particles such as colloids, proteins, and bacteria on a surface.}
Apart from proposing the algorithm with a potential distributed implementation in the network, our contribution lies in the accurate analytical characterization of the density of co-pilot users for a given total user density and a minimum distance threshold.
This result is used to characterize the probability of a pilot assignment to the typical user in the network. 

{\em 2. Two centralized pilot allocation schemes for benchmarking:} To quantify the efficacy of the proposed RSA-based pilot allocation scheme, we also propose two centralized algorithms. The first algorithm, similar to the RSA scheme, is agnostic to the RRH locations and considers only the user locations. This algorithm, named the max-min distance-based algorithm, partitions the users into sets of co-pilot users to maximize the minimum Euclidean distance among the co-pilot users. 
This scheme is optimal from the perspective of geographical separation between a set of co-pilot users.
In the proposed algorithm, the minimum distance is obtained through the bisection search subject to a set of feasibility constraints.
The second algorithm, which takes into account both RRH and user locations, maximizes the sum-user SE of the network subject to minimum user $\sinr$ constraint.
First, leveraging tools from spectral graph theory, the algorithm partitions the users into a desired number of clusters based on similar path-loss with respect to the RRHs.
Next, using the BnP algorithm, sets of co-pilot users are obtained with the additional constraint that two users in the same cluster are not assigned the same pilot. This approach provides us with a near-optimal solution in terms of sum-user SE at the cost of a significant increase in computational complexity compared to the other two algorithms. Therefore, it is more suitable to use this algorithm for benchmarking other pilot allocation schemes than practical implementation.

{\em 3: Insights from numerical results:} Through extensive system simulation, we conclude that the RSA-based pilot allocation scheme provides competitive performance compared to the max-min distance-based pilot allocation scheme, especially when the ratio of the number of users to the pilots is relatively low. Further, the RSA-based scheme achieves close to near-optimal performance with the increasing RRH density.
In addition, we compare the performance of the RSA and the max-min schemes to another centralized pilot allocation scheme based on the iterative K-means algorithm available in the literature. While RSA performs as well as the K-means, the max-min distance-based scheme marginally outperforms it.
Despite being a distributed scheme, the competitive performance of the RSA-based scheme compared to other centralized schemes makes it an attractive alternative for system implementation.

\section{System Model}\label{sec:SysMod}
\subsection{Network model} 
We limit our attention to the downlink (DL) of a cell-free mMIMO system.
The locations of the RRHs form a Poisson point process (PPP) $\Phi_r$ of density $\lambda_r$.
Similarly, the user point process $\Psi_u$ is also modeled as an independent PPP of density $\lambda_u$.
Each RRH is equipped with $N$ antennas and each user with a single antenna.
The RRHs are connected to a BBU and collectively serve users in the network.
The distance between a user at $\nbu_k \in \Psi_u$ and an RRH at $\nbr_m \in \Phi_r$ is denoted by $d_{mk}$.
In line with the mMIMO literature, where the number of antennas is assumed to be an order of magnitude more than the number of users, we consider that the antenna density $N \lambda_r \gg \lambda_u$.
Further, invoking stationarity of this setup, we analyze the system performance for the typical user $\nbu_o$, which is located at the origin $\nbo$.

{\em Channel estimation:}
Let $\nbg_{mk} = \sqrt{\beta_{mk}} \nbh_{mk}$ be the channel gain between the $m$-th RRH and the $k$-th user, where $\beta_{mk}$ captures the large-scale channel gain and $\nbh_{mk} \sim {\cal CN}(0, \nbI_N)$ captures the small-scale channel fluctuations.
We consider that the large-scale channel gain $\beta_{mk}$ is only due to the distance dependent path-loss, i.e. $\beta_{mk} =  l(d_{mk})^{-1}$, where $l(\cdot)$ is a non-decreasing  path-loss function. While the analysis presented in this paper is agnostic to the choice of $l(\cdot)$, we will need to choose a specific $l(\cdot)$ for the numerical results, which is presented in Section~\ref{sec:Results}. 

In order to obtain the channel estimates, each user uses a pilot from a set of $P$ orthogonal pilot sequences ${\cal P} = [\nbp_1, \nbp_2, \ldots, \nbp_P]^T$, where $\nbp_i$ denotes the $i$-th sequence.
The length of each pilot is $\tau_p$ symbol durations, which is less than the coherence interval.
Since we assume that the $P$ sequences are orthogonal to each other, $P \leq \tau_p$ and $\nbp_i^H\nbp_j = \tau_p \mathbf{1}(i = j)$, where $\mathbf{1}(\cdot)$ denotes the indicator function.
Due to finite number of pilots, the pilot set needs to be reused across the network.
Let the pilot used by the $k$-th user be $\nbp(k)$.
During the pilot transmission phase, the received signal matrix $Y_m \in \nbbC^{N \times \tau_p}$ at the $m$-th RRH is
\begin{align*}
Y_m = \tau_p \sum_{\nbu_k \in \Psi_{u}} \nbg_{mk}  \nbp(k)^T  + W_m, 
\end{align*}
where $\rho_p$ is the normalized transmit signal-to-noise ratio ($\snr$) of each pilot symbol and $W_m$ is an additive white Gaussian noise matrix whose elements follow ${\cal CN}(0, 1)$.
At the $m$-th RRH, the least-square estimate, $\nby_{ml} \in \nbbC^{N \times 1}$, of the channel of the users that use the $l$-th sequence is
\begin{align*}
\nby_{ml} = Y_m \nbp_l^* = \tau_p \rho_p \sum_{\nbu_k \in \Phi_{ul}}  \nbg_{mk} + W_m\nbp_l^*, 
\end{align*}
where $\Phi_{ul}$ is the set of users that use the $l$-th sequence. Further, the set of users that are assigned a pilot is defined as $\Phi_u = \cup_{k=1}^{P} \Phi_{uk}$.
Assuming $\nbu_o \in \Phi_{ul}$, the minimum-mean-squared-error (MMSE) estimate of the channel of the typical user at the $m$-th RRH is given as
\begin{align*}
\hat{\nbg}_{mo} = \nbbE[\nby_{ml} {\nbg}_{mo}^H] (\nbbE[\nby_{ml} \nby_{ml}^H])^{-1} \nby_{ml}  =  \frac{\beta_{mo}}{\sum\limits_{\nbu_k \in \Phi_{ul}} \beta_{mk} + \frac{1}{\tau_p \rho_p}} \nby_{ml} = \alpha_{mo} \nby_{ml}.\numberthis
\label{eq:MMSE_LS_Est}
\end{align*}
In this case, the error vector $\tilde{\nbg}_{mk} = \nbg_{mk} - \hat{\nbg}_{mk}$ is uncorrelated to the estimated vector.
Now the estimate and the error vectors are distributed as follows~\cite{Nayebi2017}:
\begin{align*}
\hat{\nbg}_{mo} \sim {\cal CN}\left(\boldsymbol{0}, \gamma_{mo} \nbI_N\right), \
\tilde{\nbg}_{mo} \sim {\cal CN}\left(\boldsymbol{0}, \left(\beta_{mo} - \gamma_{mo}\right)\nbI_N\right),
\end{align*}
where 
$
\gamma_{mo} = \frac{\tau_p \rho_p \beta_{mo}^2}{1 + \sum_{\nbu_k \in \Phi_{ul}} \tau_p \rho_p \beta_{mk}}.
$
From the expression of $\gamma_{mo}$, it is clear that the quality of channel estimates depend on the locations of the co-pilot users in $\Phi_{ul}$.
%%\chr{Hence, pilot assignment matters, which is the objective of this work. However, before proceeding further, in the following section we present the $\sinr$ of the users.} 

{\em DL user $\sinr$:} 
Using the channel estimates, each RRH precodes the data for all the users in the network. 
In this work, we consider conjugate beamforming precoding scheme.
Since the $m$-th RRH cannot distinguish among the channels of the users that use the $l$-th pilot, it uses the normalized direction of $\nby_{ml}$ for beamforming, i.e.
the precoding vector used to transmit data to the users that use $l$-th pilot is given as
\begin{align*}
\nbw_{ml} = {\nby_{ml}}/{\sqrt{\nbbE[\|\nby_{ml}\|^2]}} = {\hat{\nbg}_{mo}}/{\sqrt{\nbbE[\|\hat{\nbg}_{mo}\|^2]}}.
\end{align*}
Now the data transmitted by the $m$-th RRH is given as 
\begin{align*}
\nbx_m =\sqrt{\rho_d} \sum_{p=1}^P \nbw_{mp}^* \sum_{\nbu_k \in \Phi_{up}} \sqrt{\eta_{mk}} q_k,
\end{align*}
where $\eta_{mk}$ is the transmission power used by the $m$-th RRH for the $k$-th user and $q_k\sim {\cal CN}(0,1)$ is the transmit symbol of the $k$-th user.
{For each RRH, we assume the following power constraint:}
$
\nbbE[\|\nbx_m\|^2] \leq \rho_d. 
$
The symbol received at $\nbo$ (that uses the $l$-th pilot) is given as 
\begin{align*}
r_o = \sum_{\nbr_m \in \Phi_r} \nbg_{mo}^T \nbx_m + n_o = &\sqrt{\rho_d} \sum_{\nbr_m \in \Phi_r} (\hat{\nbg}_{mo}^T +  \tilde{\nbg}_{mo}^T) \frac{\hat{\nbg}_{mo}^*}{\sqrt{N \gamma_{mo}}} \sqrt{\eta_{mo}} q_o \\
& + \sqrt{\rho_d} \sum_{p=1, p \neq l}^P \sum_{\nbu_k \in \Phi_{up}} \sum_{\nbr_m \in \Phi_r} \nbg_{mo}^T \frac{\hat{\nbg}_{mp}^*}{\sqrt{N \gamma_{mp}}} \sqrt{\eta_{mk}} q_k \\
& + \sqrt{\rho_d} \sum_{\nbu_{k'} \in \Phi_{ul}' } \sum_{\nbr_m \in \Phi_r} (\hat{\nbg}_{mo}^T +  \tilde{\nbg}_{mo}^T) \frac{\hat{\nbg}_{mo}^*}{\sqrt{N \gamma_{mo}}} \sqrt{\eta_{mk'}} q_{k'} + n_o,
\end{align*}
where $\Phi_{ul}' = \Phi_{ul} \setminus {\nbu_o}$, the first term on the right hand side is the desired term, the second term corresponds to multi-user interference due to non-copilot users, and the third term is the source of interference due to pilot contamination.

\subsection{Metrics for system performance analysis}
\subsubsection{DL power control and $\sinr$ of an arbitrary user}
Since our objective is to propose a scheme to reduce pilot contamination, we focus on the operational regime where pilot contamination dominates rest of the interference terms. In the following lemma, we present the $\sinr$ expression of the typical user under the assumption that the RRHs are equipped with $N \rightarrow \infty$ antennas.
% Further, we define the key performance metrics that are used for the assessment of proposed pilot allocation scheme.
\begin{lemma}
Conditioned on $\Phi_r$ and $\Phi_u$, the asymptotic $\sinr$ of the typical user is given as 
\begin{align*}
\sinr_{o, \infty} = \frac{\left(\sum_{\nbr_m \in \Phi_r} \sqrt{\eta_{mo} \gamma_{mo}}\right)^2}{\sum_{\nbu_{k} \in \Phi_{ul}'}\left(\sum_{\nbr_m \in \Phi_r} \sqrt{\eta_{mk} \gamma_{mo}}\right)^2 }. \numberthis
\label{eq:limSinr}
\end{align*}
%where $\Phi_{ul}' = \Phi_{ul} \setminus \{\nbo\}$.
\end{lemma}
\begin{IEEEproof}
The estimated symbol at the typical user can be obtained as $
\hat{q}_o = r_o/\sqrt{N} .
$
Now, using the law of large numbers, as $N \rightarrow \infty$,  $\frac{\hat{\nbg}_{mo}^T\hat{\nbg}_{mo}^*}{N} \rightarrow \gamma_{mo}$, $\frac{\tilde{\nbg}_{mo}^T\hat{\nbg}_{mo}^*}{N} \rightarrow 0$, and $\frac{{\nbg}_{mo}^T\hat{\nbg}_{mp}^*}{N} \rightarrow 0$. 
Hence, the limiting $\sinr$ converges to~\eqref{eq:limSinr}.
\end{IEEEproof}
%%Observing the $\sinr$ expression, we make the following remark regarding equal power control scheme.
%%\begin{remark}
%%If all the RRHs transmit at the same power to each user, i.e. $\eta_{mk} = \frac{1}{K},  \forall m, \forall k$, then $\sinr_k = \frac{1}{|\Phi_{ul}|}, \forall \nbu_k \in \Phi_{ul}$.  Therefore, when pilot contamination dominates rest of the interference, the equal power control is not a suitable option.
%%\end{remark}
%The $\sinr$ largely depends on the type of power control scheme. While max-min power control has been the preferred choice for performance analysis of cell-free mMIMO systems, it requires centralized decision making at the BBU. 
In this work, we consider a distributed power control scheme~\cite{Attarifar12018} where the transmission power used by the the $m$-th RRH for the typical user at $\nbo$ is given as 
\begin{align*}
\eta_{mo} = \frac{\gamma_{mo}}{P\sum_{\nbu_k \in \Phi_{ul}} \gamma_{mk}}, \numberthis
\label{eq:pcFun}
\end{align*}
such that $\sum_{\nbu_k \in \Phi_{ul}}\eta_{mk} = 1/P$. 
We assume that the RRHs allocate equal power $1/P$  to serve each set of co-pilot users.
%{Note that since the proposed pilot assignment scheme (that we discuss in the following section) solely depends on the user locations, it can be used in conjunction with other power allocation schemes such as max-min power control~\cite{Ngo2017}.}
Now, we define the following metrics for the performance analysis:

{\em i. Pilot assignment probability of the typical user:} Since the RSA-based pilot assignment is stochastic in nature, for a given realization of user locations, a few of the users may not be assigned a pilot. Hence, pilot assignment probability to the typical user is an important metric to analyze for this scheme. Let $\{{\cal I}_o = 1\}$ be the event that the user at $\nbo$ is assigned a pilot. Then the aforementioned probability is given as 
\begin{align*}
{\cal M}_o = \nbbP[{\cal I}_o = 1] = \nbbE[\mathbf{1}({\cal I}_o = 1)], \numberthis
\end{align*}
where the expectation is taken over $\Psi_u$. In Section~\ref{sec:AssgnProb}, we present our proposed approach to characterize the above quantity along with the RSA-based pilot assignment scheme. 
This quantity can be used to get an estimate of the number of pilots necessary to satisfy target pilot assignment probability for a given user density. Note that $\ncalM_o = 1$ for the max-min distance-based and the BnP-based schemes proposed in this work.

{\em ii. Average user spectral efficiency:} It is defined as 
\begin{align*}
\overline{{\tt SE}}_{o} = & \nbbE\left[{\cal I}_o \log_2(1 + \sinr_o)\right] = \nbbE\left[\log_2(1 + \sinr_o)\big| {\cal I}_o = 1\right] \nbbP[{\cal I}_o = 1], \numberthis
\label{eq:AvgUESE}
\end{align*}
where the expectation is taken over $\Phi_r, \Psi_u$.

{\em iii. Sum-user spectral efficiency:} In contrast to the above two quantities, sum-user SE is defined for a given realization of $\Phi_r$ and $\Psi_u$ over a finite observation window $W \subset \R^2$ and given as 
\begin{align*}
\Sigma_{\tt SE} = & \sum_{\nbu_j \in \Psi_u \cap W} \ncalI_j \log_2(1 + \sinr_j), \numberthis
\end{align*}
where $\ncalI_j = 1$ for the max-min and BnP-based schemes and $\ncalI_j \in \{0, 1\}$ for the RSA-based scheme.

\section{RSA-based Pilot Allocation Scheme}\label{sec:RSA}
Before delving into the proposed RSA-based pilot assignment scheme, we find it pertinent to mention the complexity associated with pilot assignment problem that serves as a motivation to propose a heuristic algorithm.
The objective of any resource allocation algorithm is to maximize a cost (reward) function subject to certain constraints due to a limited availability of resources. 
In our case, we choose the cost function to be the sum SE. 
Hence, for a given realization of the locations of RRHs and users, our objective is to maximize the sum SE by judiciously selecting the set of co-pilot users.
Inspired by the column generation approach prevalent in the linear programming literature, the problem can be formulated in the following way.
We consider each potential set of co-pilot users as a column. 
In order to have a finite dimension for the set of feasible solutions, it is imperative to consider a finite observation window with $N_u$ users $\ncalS = \{\nbu_1, \nbu_2, \ldots, \nbu_{N_u}\}$ and $N_r$ RRHs.
Let $\ncalA$ denotes the set of all the potential co-pilot user sets avoiding the null and singleton sets.
Hence, the cardinality of $\ncalA$, which is also the total number of columns, is $2^{N_u}- N_u -1$. 
As an example, consider a set of users $\ncalA_k = \{\nbu_1, \nbu_2, \nbu_3\}$.
The corresponding column for these co-channel users is given as $\nbx_k = \begin{bmatrix}
1 & 1 & 1 & 0 & \ldots & 0 \end{bmatrix}^T \in \{0, 1\}^{N_u}$.
We define the matrix $A$, where each column corresponds to a set in $\ncalA$.
Further, the cost of a set $\ncalA_j$ (equivalently, the cost of the $j$-th column $\nbx_j$) is given as 
\begin{equation}
c(\nbx_j) = 
\begin{cases}
\sum_{\nbu_i \in \ncalA_j} \log_2\left(1 +  \Gamma_{ij}\right), & \text{if} \ \Gamma_{ij} \geq \Gamma_{\min} \ \forall \nbu_i \in \ncalA_j \\
-M, & \text{if} \ \Gamma_{ij} < \Gamma_{\min} \ \text{for any} \ \nbu_i \in \ncalA_j 
\end{cases}
\label{eq:CostFun1}
\end{equation}
where the $\sinr$ of the $i$-th user in the $j$-th set is $\Gamma_{ij} = \frac{\left(\sum_{\nbr_m \in \Phi_r} \sqrt{\eta_{mi} \gamma_{mi}}\right)^2}{\sum_{\nbu_{k} \in \{\ncalA_j \setminus \nbu_i \}}\left(\sum_{\nbr_m \in \Phi_r} \sqrt{\eta_{mk} \gamma_{mi}}\right)^2 }$, $\Gamma_{\min}$ is the minimum $\sinr$ threshold, and $M$ is a large positive number.
With this definition the set of co-pilot users not satisfying the minimum $\sinr$ threshold even for a single user is (almost) never selected.
Now, we express the optimization problem as 
\begin{subequations}
\begin{align}
\max_{\Lambda} \quad	& \sum_{s = 1}^{|\ncalA|} c(\nbx_s) \lambda_s \\ \label{eq:UEperSet}
\text{s.t.} \quad	& A \Lambda = \boldsymbol{1} \\ \label{eq:TotalPilots}
					& \|\Lambda\|_1 = P \\ \label{eq:IntConstr}
					& \Lambda \in \{0, 1\}^{|\ncalA|}, 
\end{align}
\label{eq:OptProb1}
\end{subequations}
where $\Lambda = \left[\lambda_1, \lambda_2, \ldots, \lambda_{|\ncalA|}\right]^T$, \eqref{eq:UEperSet} ensures that each user is assigned exactly one pilot, \eqref{eq:TotalPilots} ensures that $P$ columns are selected each representing a set of co-pilot users. 
Above problem is NP-hard. Further, the feasible solution space of the problem is $\tbinom{|\ncalA|}{P}$. Hence, if we wish to obtain the optimal solution even for a moderately small system of 24 users with 6 pilots, we need to search over a feasible set of size approximately $3.1 \times 10^{40}$.

Owing to the complexity of the problem, it is natural to consider heuristic solutions, {\em albeit} sub-optimal, that can be implemented efficiently in the network. 
In the following subsection, we present a sub-optimal pilot allocation algorithm that only considers user locations to select the set of co-pilot users such that the pilot contamination-based interference is mitigated thereby implicitly improving the user SE as well as the sum SE. This algorithm, which is inspired by the RSA process, can be implemented both in a centralized or distributed manner and is easily scalable as the network size grows. 
%%In order to quantify the efficacy of this algorithm, we propose a second centralized scheme that is based on solving a optimization problem such that the minimum distance between a set of copilot users is maximized.
%%Next, we begin our discussion with the RSA-inspired pilot allocation scheme.

\subsection{RSA-based pilot assignment algorithm}\label{sec:AssgnProb}
{Our goal is to select the sets of co-pilot users among all the users in the network such that a minimum distance $R_{\tt inh}$ is maintained between two co-pilot users.
This can be achieved by dependent selection of the users from the original user point process $\Psi_u$ as outlined in Algorithm~\ref{algo:PilotAssgnAlgo}.} 
The algorithm assigns a random mark $t_k$, which is uniformly distributed in $[0, 1]$, to each point $\nbu_k \in \Psi_u$.
Let ${\cal B}_{R_{\tt inh}}(\nbu_k)$, a circle of radius $R_{\tt inh}$ centered at $\nbu_k$, be defined as the contention domain of the point at $\nbu_k$.
For pilot assignment, the algorithm considers each user in increasing order of their marks, i.e. the lowest mark is considered first.
From the available set of pilots, a pilot is randomly assigned to a user at $\nbu_k$, where the set of available pilots are those which have not been assigned to the users in ${\cal B}_{R_{\tt inh}}(\nbu_k)$.
Note that to implement this algorithm, the BBU requires only the location information of the users, which does not require any additional signaling overhead as this information is typically present at a centralized node in the network such as the BBU. 
At the end of this subsection, we also discuss a protocol for potential distributed implementation of the algorithm.

\begin{algorithm}
\hrule
\vspace{0.05cm}
{\small
 \KwInput{User locations $\Psi_u$, the set of pilots ${\cal P}$, inhibition distance $R_{\tt inh}$}\vspace{-0.25cm}
 \KwResult{Pilot assignment table ${\cal T}$}\vspace{-0.25cm}
 Initialization: ${\cal T} = \emptyset$, a random mark $t_i \sim {U}(0, 1)$ for each $\nbu_i \in \Psi_u$\;\vspace{-0.25cm}
Let $\tilde{\Psi}_u$ be the set of users in the increasing order of marks\;\vspace{-0.25cm}
 \For{User $\nbu \in \tilde{\Psi}_u$}{\vspace{-0.25cm}
  Set: ${\cal P'} = {\cal P}$\;\vspace{-0.25cm}
  \While{User $\nbu$ is not assigned a pilot}{\vspace{-0.25cm}
  \eIf{${\cal P'} == \emptyset$}{\vspace{-0.25cm}
  	No pilot can be assigned: ${\cal T} = {\cal T} \cup \emptyset$\;\vspace{-0.25cm}
  	{\tt Break;}\vspace{-0.25cm}
  	} 
  	{ Select a pilot sequence $\nbp_k$ randomly from the set ${\cal P'}$\;\vspace{-0.25cm}
   }\vspace{-0.25cm}
   \eIf{No other users in ${\cal B}_{R_{\tt inh}}(\nbu)$ are using $\nbp_k$}{\vspace{-0.25cm}
  	Assign the pilot: ${\cal T} = {\cal T} \cup {\nbp_k}$\;\vspace{-0.25cm}
  	{\tt Break}\;\vspace{-0.25cm}
  	}{Remove $\nbp_k$ from list of potential pilots: ${\cal P'} = {\cal P'} \setminus \nbp_k$\;}\vspace{-0.25cm}
   }\vspace{-0.25cm}
 }\vspace{-0.25cm}
 }\vspace{-0.25cm}
 \hrulefill
 \vspace{0.1 cm}
\caption{The RSA-based pilot assignment algorithm in for a cell-free mMIMO system.}
\label{algo:PilotAssgnAlgo}
\end{algorithm}

\begin{figure*}[!htb]
\centering
\begin{subfigure}{0.32\textwidth}
  \centering
  \includegraphics[width=0.7\linewidth]{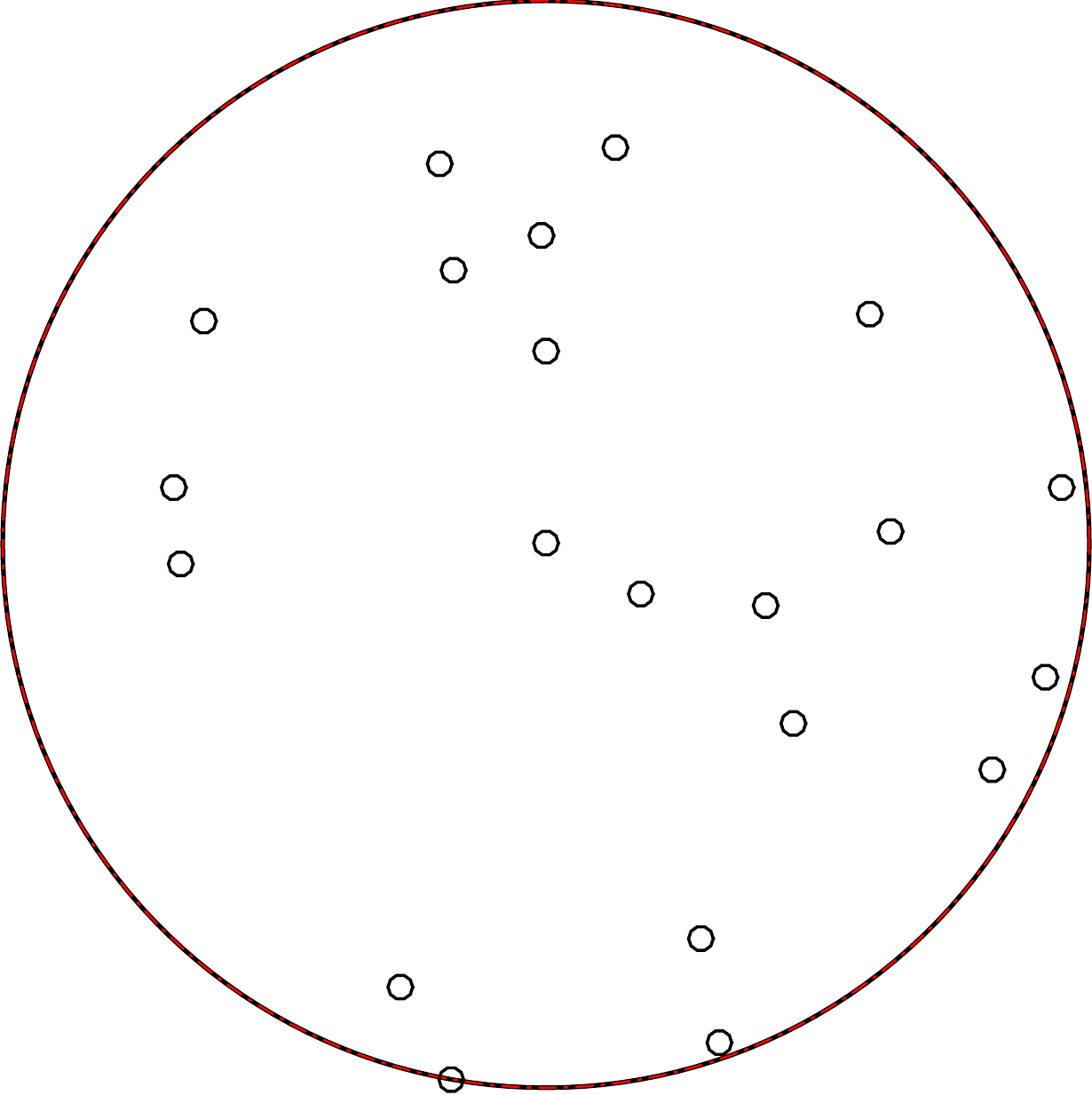}
\end{subfigure}
\begin{subfigure}{0.32\textwidth}
  \centering
  \includegraphics[width=0.7\linewidth]{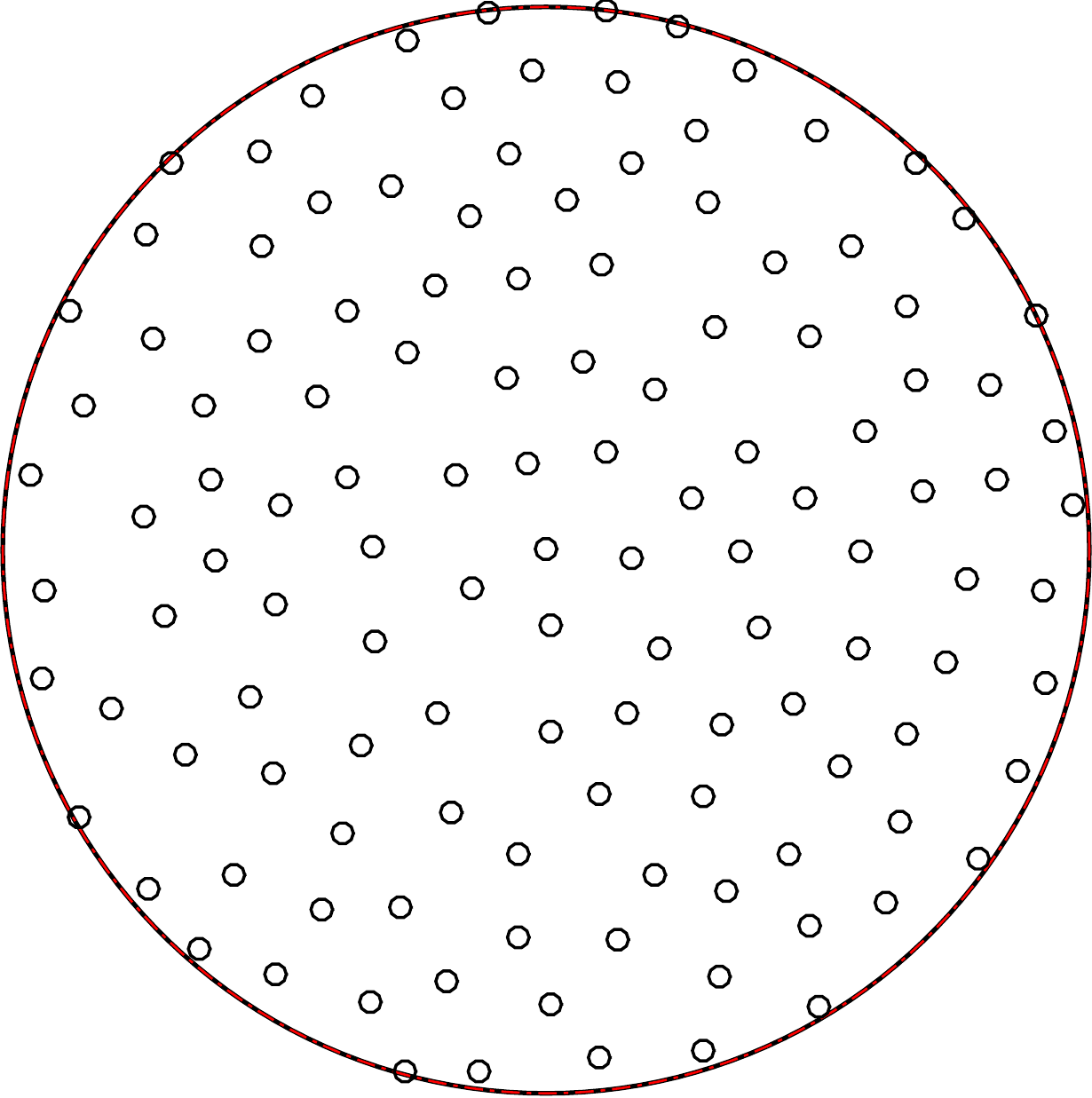}
\end{subfigure}%
\begin{subfigure}{0.32\textwidth}
  \centering
  \includegraphics[width=0.7\linewidth]{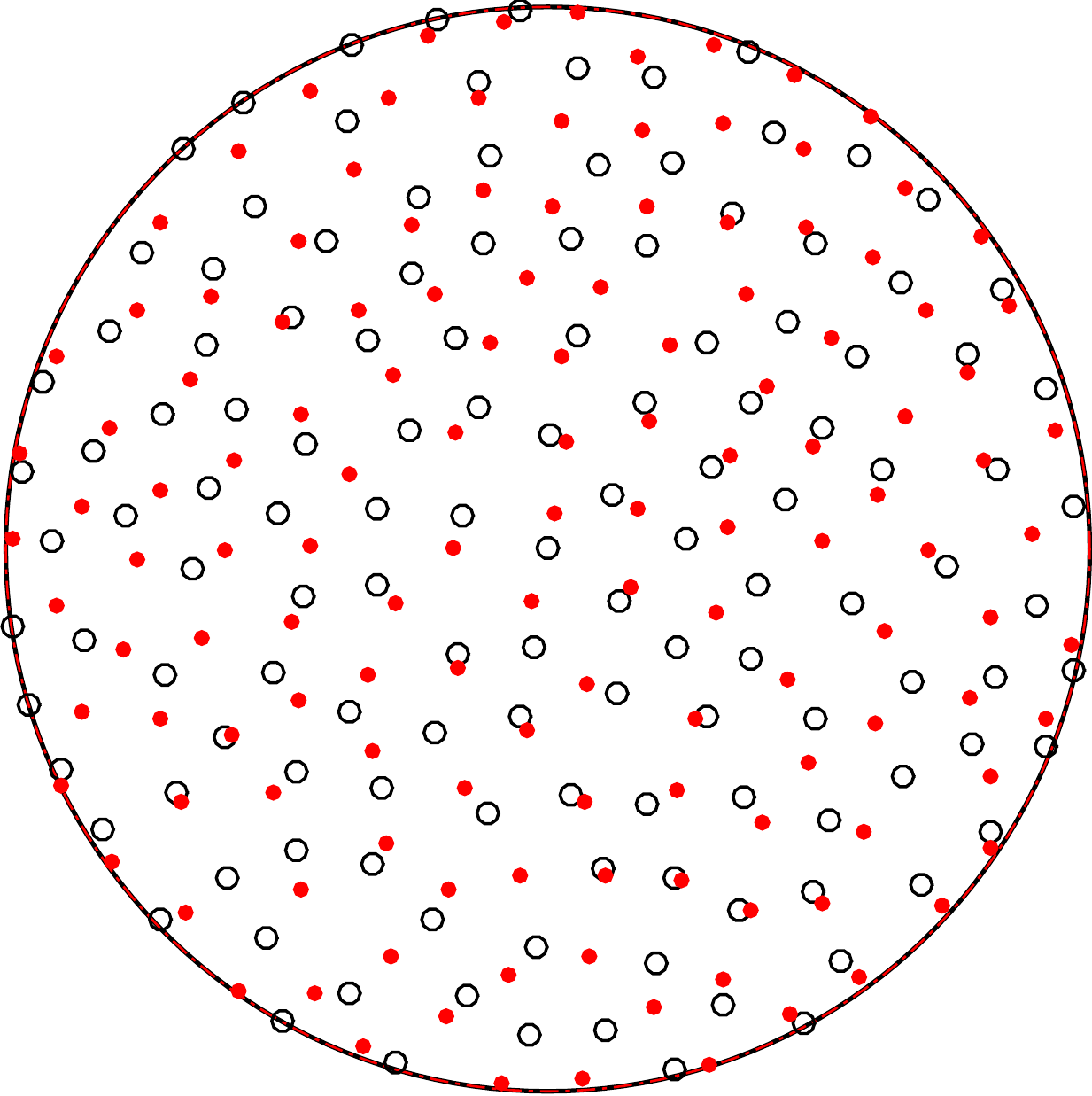}
\end{subfigure}%
\caption{\footnotesize Realizations of co-pilot user locations using Algorithm~\ref{algo:PilotAssgnAlgo}. Parameters: $R_{\tt inh} = 200$, $\lambda_u = 2 \times 10^{-6}$ (left), $\lambda_u = 10^{-3}$ (center, right). Left and center figures represent realizations of co-pilot users for $P=1$. Right figure represents a realization of co-pilot users for $P=2$.}
\label{fig:CoPilotRealization}
\end{figure*}

For the system designers, it is useful to know the probability that a user will be scheduled as a function of the density of users and the number of pilots in the system. Following subsections present an approximate theoretical result that answers the aforementioned question eliminating the need for a system simulation. It is worth-mentioning that the approximate result is a new contribution to the RSA literature as the exact solution for counterpart of this problem even in the case of 1D is unknown.

\subsubsection{Analysis of the pilot assignment probability}
Recall that $\Phi_u$ is the set of users that are assigned a pilot (in this case by Algorithm~\ref{algo:PilotAssgnAlgo}), 
i.e. $\Phi_u = \cup_{p=1}^P \Phi_{up}$.
%Based on the algorithm, each user in $\Psi_u$ is assigned a mark, which is uniformly distributed in $[0, 1]$.
Let $t_o$ be the mark associated with the typical user.
%For each user $\nbu_k \in \Psi_u$, we define its contention domain as ${\cal B}_{R_{\tt inh}}(\nbu_k)$.
Now, the user at $\nbo$ is assigned a pilot if $|\Phi_u \cap {\cal B}_{R_{\tt inh}}(\nbo)| \leq P-1$. This is ensured by the following two events: 
\begin{itemize}
\item ${\cal E}_1$: there are at most $P-1$ points in ${\cal B}_{R_{\tt inh}}(\nbo) \cap \Psi_u$ that have marks less than $t_o$, 
\item ${\cal E}_2$: there are more than $P-1$ points in ${\cal B}_{R_{\tt inh}}(\nbo) \cap \Psi_u$ that have marks less than $t_o$. However, some of these points are not assigned a pilot as their contention domains have more than $P$ points with marks smaller than their respective marks.
\end{itemize}
While obtaining the probability of ${\cal E}_1$ is straightforward, characterizing ${\cal E}_2$ is highly non-trivial even for $P=1$.
{Note that for $P = 1$, the above formulation has been used to model the CSMA-CA networks. However, due to the intractability of ${\cal E}_2$,  Mat\'{e}rn hardcore process of type-II (MHPP-II) has been used for approximate characterization for $\Phi_u$~\cite{NguBac2007}. Hence, one may be inclined to extend the MHPP-II process for $P \geq 2$. However, one of the limitations of the MHPP-II process is that it underestimates the number of points in $\Phi_u$~\cite{Busson2014}. 
Hence, the extension of the MHPP-II model for $P \geq 2$ will not result in an accurate estimation.}
On the other hand, for $P=1$, $\Phi_u$ is exactly modeled by the simple sequential inhibition (SSI) process~\cite{Busson2014} or the RSA process~\cite{Talbot2000}. 
Using this fact, in the sequel, we present an efficient heuristic to estimate the pilot assignment probability.

Consider a finite observation window ${\cal B}_{R_s}(\nbo) \subset \R^2$, where $R_s \gg R_{\tt inh}$. 
Let $N_u = |\Psi_u \cap {\cal B}_{R_s}(\nbo)|$ be the total number of users and $N_s = |\Phi_u \cap {\cal B}_{R_s}(\nbo)|$ be the number of users that are assigned a pilot. 
Note that for a given $N_u$, $N_s$ is a random variable as it depends on the realization of  $\Psi_u$ as well as random marks associated with these points.
Now, for a given $N_u > P$, $\nbbE[N_s|N_u]$ is the average number of users that are assigned a pilot. Hence, the probability that the typical user out of the $N_u$ users is assigned a pilot is $\nbbE[N_s|N_u]/N_u$. 
On the other hand, for $N_u \leq P$, the typical user is assigned a pilot with probability 1. 
Combining these two events, we write the pilot assignment probability as 
\begin{align*}
\nbbP[{\cal I}_o = 1] = &\nbbP\left[N_u \leq P\right] +  \nbbE_{N_u}[{\nbbE[N_s|N_u]}N_u^{-1}\big|N_u > P] \nbbP\left[N_u > P\right] \\ 
\approx & \nbbP\left[N_u \leq P\right] + \nbbE\left[N_s | N_u = \pi R_s^2 \lambda_u\right] \nbbE[N_u^{-1}\big|N_u > P] \nbbP\left[N_u > P\right], \numberthis
\label{eq:PAProb}
\end{align*}
where the second step is an approximation as instead of $\nbbE_{N_u}[\nbbE[N_s|N_u]|N_u > P]$, we determine $\nbbE[N_s]$ by considering $N_u = \pi R_s^2 \lambda_u$, which is its expected value.
Since $N_u$ is Poisson distributed with mean $\lambda_u \pi R_s^2$, 
\begin{align*}
\nbbE\left[N_u^{-1}|N_u > P\right] = \frac{\lambda_u \pi R_s^2 - \sum_{n=0}^{P} \text{Poi}(\lambda_u \pi R_s^2, n)}{1- \sum_{n=0}^{P}\text{Poi}(\lambda_u \pi R_s^2, n)},\numberthis
\label{eq:N_uInverse}
\end{align*}
where $\text{Poi}(\lambda_u \pi R_s^2, n) = e^{-\lambda_u \pi R_s^2}\frac{(\lambda_u \pi R_s^2)^n}{n!}$.

Next, we discuss our approach to analytically estimate $\nbbE\left[N_s | N_u = \pi R_s^2 \lambda_u \right]$ leveraging the rich theory of the RSA process.
For convenience, we use the notation $\nbbE\left[N_s\right]$ to represent the above expectation. 
We first present the analysis for the special case of $P=1$ followed by its extension to the general case of $P \geq 2$. 

\subsubsection{Pilot assignment probability for $P = 1$}
Traditionally, the RSA process has been used across different disciplines, such as condensed matter physics, surface chemistry, and cellular biology, to study the adsorption of different substances, such as colloids, proteins, and bacteria, on a surface~\cite{Talbot2000}.
Next, we present a brief overview of the RSA process before analyzing pilot assignment probability.

{\em Random sequential adsorption process:}
An RSA process is defined as a stochastic space-time process, where $n$-dimensional hard spheres sequentially arrive at random locations  in $\R^n$ such that any arriving sphere cannot overlap with already existing sphere. 
More formally, for 2D case, let $\Psi$ be a homogeneous space-time point process on $\R^2 \times \R^+$. 
The circles with radii $R_{\tt inh}/2$ are arriving at a rate of $\lambda_{\Psi}$ per unit area. 
Let $\Psi(t)$ be the point process on $\R^2$ when $\Psi$ is observed at an arbitrary time $t$.
Observe that the density of $\Psi(t)$ is $\lambda_{\Psi}t$.
At time $t$, an arriving point at $\nbx \in \R^2$ is retained if there are no other points within ${\cal B}_{R_{\tt inh}}(\nbx)$.
{Let $\varphi(t|\Psi)$ be a realization of the set of the retained points at time $t$.
Clearly, $\varphi(t_1|\Psi) \subseteq \varphi(t_2|\Psi)$ for $t_1 \leq t_2$. Moreover, there exists a time $c \in R^+$ such that $\varphi(t_i|\Psi) = \varphi(t_j|\Psi)$ for $t_i, t_j > c$, i.e. no more points can be added to the system.}
This is known as the {\em jamming limit}.
Observe that the random marks assigned by Algorithm~\ref{algo:PilotAssgnAlgo} can be thought of as the arrival times of the points in $\Psi_u$.
In this interpretation, the points that arrive early (have smaller marks) are more likely to get an assignment.

{Let $\Phi(t)$ be the point process of the retained points at time $t$ and $\rho(t)$ be the corresponding density.}
In order to obtain the density of retained point process for a given density of original point process, we need to observe the system at a specific time. 
For example, if we want to obtain the density of retained points for an original point density of $2 \lambda_{\Psi}$, then we need to observe the system at $t = 2$.
Fig.~\ref{fig:CoPilotRealization} illustrates the realizations of co-pilot users for different $\lambda_u$ and $P$.
In the left figure, the system does not reach the jamming limit due to lower density of the original user point process $\Psi_u$. 
On the other hand, the center figure (almost) reaches the jamming limit and there cannot be more co-pilot users in the system. 
Notice the regular, almost grid-type, realization of points.
The right figure illustrates the jamming state for $P = 2$. 
In the following lemma, we present the density of $\Phi(t)$.

\begin{lemma}\label{lem:RSA_Density}
The density $\rho(t)$ of the point process $\Phi(t)$ is obtained by solving the following differential equation~\cite{schaaf1989} with the initial condition $\rho(0) = 0$:
\begin{align*}
\int \frac{{\rm d}\rho(t)}{\phi(\kappa \rho(t))} = \frac{\lambda_{\Psi}}{\kappa} t + C, \numberthis
\label{eq:DiffEqun}
\end{align*}
where $\kappa = \frac{\pi R_{\tt inh}^2}{4}$ is the area covered by a circle, $\kappa \rho(t)$ is the fraction of the area that is covered by the retained circles at time $t$, $\phi(\kappa \rho(t))$ is the probability that a circle arriving at an arbitrary location in $\R^2$ is retained at time $t$, and $C$ is the integration constant.
The retention probability is given as~\cite[Eq.~19]{schaaf1989} $\phi(\kappa \rho(t)) = $
\begin{align*}
& 1 - 4 \pi R_{\tt inh}^2 \rho(t) + \frac{\rho(t)^2}{2} \int\limits_{R_{\tt inh}}^{2 R_{\tt inh}} 4 \pi r A_2(r) {\rm d}r + \frac{\rho(t)^3}{3} \int\limits_{R_{\tt inh}}^{2 R_{\tt inh}} 2 \pi r A_2^2(r) {\rm d}r - S_3^{\tt eq} + O(\rho(t)^4), \numberthis
\label{eq:ExclusionProb}
\end{align*}
where $S_3^{\tt eq} = \frac{\rho(t)^3}{8} \pi \left(\sqrt{3} \pi - \frac{14}{3}\right)R_{\tt inh}^6$, $A_2(r)$ is the area of intersection of two circles of radius $R_{\tt inh}$ whose centers are separated by distance $r$.
\end{lemma}
\begin{IEEEproof}
For the detailed proof of this lemma, please refer to~\cite{schaaf1989}. 
Due to space limitations, we just present the proof sketch here. 
Note that $\kappa\rho(t)$ is the fraction of area covered by the retained circles at time $t$.
Now, the rate of change of the fraction of the covered area with respect to time depends on the number of arrivals $\lambda_{\Psi}{\rm d}t$ per unit area and the probability of an arrival being retained, which is given by $\phi\left(\kappa \rho(t)\right)$. Hence, 
\begin{align*}
 \frac{{\rm d} (\kappa\rho(t))}{{\rm d}t} = \lambda_{\Psi} \phi\left(\kappa \rho(t)\right). \numberthis
\label{eq:Langmuir_RSA}
\end{align*}
The expression for $\phi\left(\kappa \rho(t)\right)$ is derived in~\cite{schaaf1989}.
Solution of the differential equation \eqref{eq:Langmuir_RSA} gives the density of $\Phi(t)$.
\end{IEEEproof}

Since the function~\eqref{eq:ExclusionProb} is difficult to work with, a fitting function is analytically presented in \cite{schaaf1989} as $\phi_{\tt FIT}(\rho(t)) =$
\begin{align*}
 (1 + b_1 x(t) + b_2 x(t)^2 + b_3 x(t)^3)(1-x(t)^3), \numberthis
\label{eq:FitFun}
\end{align*}
where $x(t) = \rho(t)/\rho(\infty)$ and $\rho(\infty)\kappa = 0.5474$ is the fraction of the area that is covered at the jamming limit as $t \rightarrow \infty$. The coefficients $b_1, b_2$ and $b_3$ are obtained by matching the order of $\rho(t)$ in equations~\eqref{eq:ExclusionProb} and \eqref{eq:FitFun}.
Now the expression for $\rho(t)$ is obtained by solving the differential equation \eqref{eq:DiffEqun}. 
While the closed form solution of the equation is difficult, the problem can be efficiently solved using standard numerical softwares.
Now, with the help of Lemma~\ref{lem:RSA_Density}, we present pilot assignment probability to a user for $P = 1$.
\begin{lemma}\label{lem:PAP1}
For a system with $P = 1$, the probability that the typical user is assigned a pilot is
\begin{align*}
\nbbP[{\cal I}_o = 1] \approx (1 + \pi R_s^2 \lambda_u)e^{-\pi R_s^2 \lambda_u}  + (1 - (1 + \pi R_s^2 \lambda_u)e^{-\pi R_s^2 \lambda_u}) (\rho(1)\pi R_s^2) \nbbE\left[N_u^{-1}| N_u > 1\right],
\end{align*}
where $\rho(1)$ is determined using Lemma~\ref{lem:RSA_Density} and $\nbbE\left[N_u^{-1}| N_u > 1\right]$ using \eqref{eq:N_uInverse}.
\end{lemma}
\begin{IEEEproof}
Since the density of user process $\Psi_u$ is $\lambda_u$ users per unit area, as per the RSA process definition, we can construct an equivalent space-time process where the arrivals occur at $\lambda_u$ users per unit area per unit time. Now, to obtain the density of $\Phi_u$, we observe this space-time system at time $t = 1$. Hence, the density of $\Phi_u$ is $\rho(1)$.
The final expression is obtained by replacing $\nbbE[N_s] = \pi R_s^2 \rho(1)$ in \eqref{eq:PAProb}.
\end{IEEEproof}

%%\chr{Moreover, in the following figure, we present the density of co-pilot users as a function of the density of initial user density $\lambda_u$.
%%As evident from Fig.~\ref{fig:lamSSI_vs_lamPPP}, the co-pilot user density as a function of initial user density $\lambda_u$ can be divided into two regimes: (1) unsaturated regime, (2) saturated regime.
%%Depending on $R_{\tt inh}$, the saturated regime is reached at different $\lambda_u$s.}

\subsubsection{Pilot assignment probability for $P \geq 2$}

For the general case of $P \geq 2$, consider that $\Phi_{u1}, \Phi_{u2}, \ldots, \Phi_{uP}$ contain the locations of the users that are assigned the pilots $\nbp_1, \nbp_2, \ldots, \nbp_P$, respectively, by Algorithm~\ref{algo:PilotAssgnAlgo}.
Since Algorithm~\ref{algo:PilotAssgnAlgo} has no preference regarding the pilots, the densities of $\Phi_{u1}, \Phi_{u2}, \ldots, \Phi_{uP}$ are the same. Let ${\lambda}_{\Phi_{uo}}$ be this density.
In order to determine ${\lambda}_{\Phi_{uo}}$, modifications in Lemma~\ref{lem:RSA_Density} are necessary.
To be specific, for \eqref{eq:ExclusionProb}, the knowledge of virial coefficients for a mixture of non-interacting hard spheres, and subsequently derivation of $S_3^{\tt eq}$ is necessary~\cite{schaaf1989}.
Since the above steps appear extremely difficult for this case, we provide an approximate yet accurate way to estimate the pilot assignment probability for $P \geq 2$.

\begin{algorithm}{\small
\hrule \vspace{0.05cm}
 \KwInput{User locations $\Psi_u$, the set of pilots ${\cal P}$, inhibition distance $R_{\tt inh}$;}
 \KwResult{Pilot assignment table ${\cal T}$;}
 Initialization: $\Psi_u' = \Psi_u$, ${\cal T} = \emptyset$\;
 \For{Each pilot $\nbp_k \in {\cal P}$}{
 \For{Each user $\nbu \in \Psi_u'$}{
  \If{No other users in ${\cal B}_{R_{\tt inh}}(\nbu)$ are using $\nbp_k$}{
  	Assign the pilot: ${\cal T} = {\cal T} \cup {\nbp_k}$\;
  	Remove $\nbu$ from list of users: $\Psi_u' = \Psi_u' \setminus \nbu$\;
  	}
   }
 }}
\hrulefill
\vspace{0.05cm}
\caption{The regenerative algorithm for pilot assignment.}
\label{algo:Regenerative}
\end{algorithm}

First, we present the regenerative pilot assignment algorithm (Algorithm~\ref{algo:Regenerative}) that is essential for our approximate analysis.
Different from Algorithm~\ref{algo:PilotAssgnAlgo}, in Algorithm~\ref{algo:Regenerative}, the pilots are assigned to users sequentially, i.e. for the typical user the second pilot sequence is considered if the first pilot has already been assigned to a user in its contention domain, the third pilot sequence is considered if both the first and the second pilots have been used in its contention domain, and so on. 
%{Recall that at the jamming limit, no additional co-pilot user can be added to the system.}
In order to proceed with our analysis, we make the following remark:
\begin{remark}\label{rem:SameDen}
The total number of pilot reuses required in ${\cal B}_{R_{s}}(\nbo)$ to obtain a target pilot assignment probability is the same for both Algorithms~\ref{algo:PilotAssgnAlgo} and \ref{algo:Regenerative}.
In other words, the density of users that are assigned a pilot is the same for both the algorithms.
\end{remark}

Let $\tilde{\Phi}_{u1}, \tilde{\Phi}_{u2}, \ldots, \tilde{\Phi}_{uP}$ contain the locations of the users that are assigned pilots $\nbp_1, \nbp_2, \ldots, \nbp_P$, respectively, by Algorithm~\ref{algo:Regenerative}.
Let ${\lambda}_{\tilde{\Phi}_{u1}}, {\lambda}_{\tilde{\Phi}_{u2}}, \ldots, {\lambda}_{\tilde{\Phi}_{uP}}$ be the densities of $\tilde{\Phi}_{u1}, \tilde{\Phi}_{u2}, \ldots, \tilde{\Phi}_{uP}$, respectively.
We obtain these densities by sequentially using Lemma~\ref{lem:RSA_Density}.
First, the density ${\lambda}_{\tilde{\Phi}_{u1}}$ of the users that are assigned the pilot $\nbp_1$ is directly obtained from Lemma~\ref{lem:RSA_Density} where the initial density of the process is $\lambda_u$.
Now, to obtain the density ${\lambda}_{\tilde{\Phi}_{u2}}$ of the users that are assigned the pilot $\nbp_2$, we approximate the initial density of users as $\lambda_u - {\lambda}_{\tilde{\Phi}_{u1}}$.
Also note that the points in $\Psi_u \setminus \tilde{\Phi}_{u1}$ do not form a PPP. However, for simplicity we approximate $\Psi_u \setminus \tilde{\Phi}_{u1}$ as a PPP.
Similarly, to obtain ${\lambda}_{\tilde{\Phi}_{u3}}$, we approximate $\Psi_u \setminus \{\tilde{\Phi}_{u1} \cup \tilde{\Phi}_{u2}\}$ as a PPP of density $\lambda_u - {\lambda}_{\tilde{\Phi}_{u1}} - {\lambda}_{\tilde{\Phi}_{u2}}$ and use Lemma~2.
The same approximation is made to get the rest of the densities.
In the next section, we will demonstrate that these approximations do not compromise the accuracy of our results.
Based on Remark~\ref{rem:SameDen}, with the knowledge of ${\lambda}_{\tilde{\Phi}_{u1}}, {\lambda}_{\tilde{\Phi}_{u2}}, \ldots, {\lambda}_{\tilde{\Phi}_{uP}}$, we can obtain
\begin{align*}
\lambda_{\Phi_{uo}} = {\sum_{l=1}^{P} \lambda_{\tilde{\Phi}_{ul}}}/{P}.\numberthis
\label{eq:MultiRSA_Density}
\end{align*}
In the next lemma, we present the pilot assignment probability for the general case of $P \geq 1$.
\begin{lemma}\label{lem:PAPgeq1}
For a system with $P \geq 1$, the pilot assignment probability for the typical user is 
\begin{align*}\label{eq:PAmultiRSA}
\nbbP[{\cal I}_o = 1] \approx \nbbP[N_u \leq P] + \nbbP[N_u > P] (P \lambda_{\Phi_{uo}} \pi R_s^2) \nbbE[N_u^{-1}| N_u > P],
\numberthis
\end{align*}
where $\lambda_{\Phi_{uo}}$ is determined from \eqref{eq:MultiRSA_Density} and Lemma~\ref{lem:RSA_Density}, $\nbbE[N_u^{-1}| N_u > P]$ is determined using \eqref{eq:N_uInverse}, and $N_u$ is Poisson distributed with mean $\lambda_u \pi R_s^2$.
\end{lemma}
\begin{IEEEproof}
The proof follows on the similar lines as that of Lemma~\ref{lem:PAP1}.
\end{IEEEproof}

\subsection{Distributed implementation of the RSA-based pilot allocation scheme}
The RSA based pilot allocation scheme can also be implemented in a distributed manner. 
Consider the moment when a user $\nbu_o$ enters the network. 
During the initial access phase, the user senses the environment to get an estimate of active pilot transmission in the vicinity. 
Let $\nbr_o \in \nbbC^{1 \times \tau_{\tt IA}}$ be the received signal  obtained through sensing. 
Note that $\tau_{\tt IA}$ should span over multiple coherence time intervals $\tau_c$ to average out the effect of small scale fading. 
Assuming synchronization has been established between the network and the user, the received signal strength on $k$-th pilot can be estimated as 
\begin{equation}
P_k = \frac{1}{\tau_{\tt IA}/\tau_c} \sum_{m=1}^{\tau_{\tt IA}/\tau_c} \nbr_o[(m-1)\tau_c+1:(m-1)\tau_c + \tau_p]^H \nbp_k,
\label{eq:PowPerPilot}
\end{equation}
where $\tau_p$ is the duration of the pilot sequence. 
Once the received signal powers on all the pilots are calculated, they are compared with a threshold power $P_{\tt inh}$, which is a function of $R_{\tt inh}$.
A pilot is randomly selected from the set of candidate pilots $\{\nbp_k : P_k \leq P_{\tt inh}\}$.
If this set is empty, then the user is not assigned a pilot. Algorithm~\ref{algo:DistributedRSA} presents the above-mentioned procedure.

\begin{algorithm}{\small
\hrule \vspace{0.05cm}
 \KwInput{Power threshold $P_{\tt inh}$, Received signal $\nbr_o$;} \vspace{-0.25cm}
 \KwResult{Pilot for user $\nbu_o$;}\vspace{-0.25cm}
  Initialization: Candidate set of pilots $\ncalP_c = \emptyset$ \;\vspace{-0.25cm}
 \For{Each pilot $\nbp_k \in {\cal P}$}{ \vspace{-0.25cm}
  	Obtain $P_k$ using \eqref{eq:PowPerPilot} \; \vspace{-0.25cm}
  	\If{$P_k \leq P_{\tt inh}$}{ \vspace{-0.25cm}
	  	$\ncalP_c = \ncalP_c \cup \nbp_k$ \; \vspace{-0.25cm}
  	}\vspace{-0.25cm}  	
 }\vspace{-0.25cm}
 \If{$\ncalP_c \neq \emptyset$}{
 Select a pilot randomly from $\ncalP_c$.\vspace{-0.25cm}
 } \vspace{-0.25cm}
 }\vspace{-0.25cm}
\hrulefill
\vspace{0.05cm}
\caption{The algorithm for an arriving user to select a pilot during initial access phase.}
\label{algo:DistributedRSA}
\end{algorithm}

\vspace{-0.5cm}

\section{Max-min Distance-based Pilot Allocation Scheme}
While the proposed RSA-based scheme is a computationally efficient scheme with possible distributed implementation, a natural question is how good is the quality of the solution.
In this section, we propose an algorithm that has the objective of maximizing the minimum distance between the set of co-pilot users similar to the RSA-based scheme.
However, in contrast to the RSA scheme that can be implemented in a distributed manner, this algorithm can only be implemented in a centralized way and does not have the scalability property of the RSA-based scheme.

To have a meaningful problem formulation, we restrict our attention to a finite spatial observation window $W \subset \R^2$. 
Let the set of users in this observation window be given as $\ncalS = \{\nbu_1, \nbu_2, \ldots, \nbu_{N_u}\}$. 
Our objective is to partition $\ncalS$ into $P$ sets $\ncalS_1, \ncalS_2, \ldots, \ncalS_P$ such that the minimum distance between any two users in a partition is maximized.
For a user $\nbu_n$, the binary variable $y_{nk} = 1$ if the user belongs to $\ncalS_k$ and $0$ otherwise.
We define the metric $d_{\min}(\ncalS_k) := \min\{\|\nbu_i - \nbu_j\| : \nbu_i \neq \nbu_j, y_{ik}=y_{jk} = 1 \}$ as the minimum distance between two elements in $\ncalS_k$.
The problem of maximizing the minimum distance between users belonging to the same set can be written as 
\begin{subequations}
\begin{align}
\max_{\{y_{nk}\}} \min_{k = 1, \ldots, P}  \quad	&  d_{\min}(\ncalS_k) \\ \label{eq:MMP1C1}
\text{s.t.} \quad	& \sum_{k=1}^P y_{nk} = 1, \quad \forall n = 1, 2, \ldots, N_u, \\ \label{eq:MMP1C2}
					& \sum_{n=1}^{N_u} y_{nk} > 1, \quad \forall k = 1, 2, \ldots, P, \\ \label{eq:MMP1C3}
					& y_{nk} \in \{0, 1\}, \quad \forall n,  \forall k, 
\end{align}
\label{eq:MaxMinProb1}
\end{subequations}
where \eqref{eq:MMP1C1} ensures that each point belongs to exactly one partition, \eqref{eq:MMP1C2} ensures that each partition has more than two points, \eqref{eq:MMP1C3} imposes the integrality constraint. 
Note that \eqref{eq:MMP1C2} can be modified to ensure more balanced partitioning. For example, if we need each partition to have more than $x \leq N_u/P$ users then we can set the constraint as $\sum_{n=1}^N y_{nk} > x, \quad \forall k$.
This problem can be reformulated as 
\begin{subequations}
\begin{align}
\max_{t, y_{nk}} \quad t \\ \label{eq:MMP2C1}
\text{s.t.} \quad	&  \|\nbu_i-\nbu_j\|  > t y_{ik} y_{jk}  \quad \nbu_i \in \ncalS, \nbu_j \in \ncalS,  i \neq j, k = 1, 2, \ldots P, \\ 
					& \sum_{k=1}^P y_{nk} = 1, \quad \forall n = 1, 2, \ldots, N_u, \\ 
					& \sum_{n=1}^{N_u} y_{nk} > 1, \quad \forall k = 1, 2, \ldots, P, \\ 
					& y_{nk} \in \{0, 1\}, \quad \forall n,  \forall k,
\end{align}
\label{eq:MaxMinProb2}
\end{subequations}
where \eqref{eq:MMP2C1} ensures that two points belonging to the same partition are separated by distance $t$ and rest of the constraints are the same as the previous formulation. The aforementioned problem can be solved in two steps. In the first step, a bisection search is used to improve the objective function, and in the second step for a given $t$, a feasibility problem is solved. 
The optimization routine to solve the problem is presented in Algorithm~\ref{algo:MaxMin}.

\begin{algorithm}{\small
\hrule \vspace{0.05cm}
\begin{algorithmic}[1]
\STATE {\em Initialization:} Set the values of $t_{\min}$ and $t_{\max}$ that define the solution space for the bisection search. Select a tolerance parameter $\epsilon$. 

\STATE Set $t = (t_{\min} + t_{\max})/2$. Solve the following feasibility problem:
\begin{subequations}
\begin{align}
&   \sum_{k=1}^K y_{nk} = 1 \quad \forall n = 1, 2, \ldots, N_u \\ 
& \sum_{n=1}^{N_u} y_{nk} > 1 \quad \forall k = 1, 2, \ldots, P \\ 
& y_{ik} + y_{jk} \leq 1 \quad \forall \|\nbu_i - \nbu_j\| < t, \\
& y_{nk} \in \{0, 1\} \quad \forall n,  \forall k.
\end{align}
\label{eq:FeasibilityProb}
\end{subequations}\vspace{-0.25cm}
\IF{\eqref{eq:FeasibilityProb} is feasible} \STATE Set $t_{\min} = t$. 
\ELSE
	\STATE Set $t_{\max} = t$.
\ENDIF
\STATE Repeat the above steps until if $|t_{\max} - t_{\min}| < \epsilon$. 
\end{algorithmic}}\vspace{-0.25cm}
\hrulefill
\vspace{0.1cm}
\caption{Solving the max-min distance partitioning problem}
\label{algo:MaxMin}
\end{algorithm}

%%\subsection{$K$-means clustering based pilot allocation}
%%In order to assign pilot in a cell-free network, an iterative $K$-means clustering-based pilot allocation is proposed in~\cite{Attarifar12018}.
%%The algorithm is presented in  Algorithm~\ref{algo:kmeans} for the sake of completeness. 
%%The algorithm operates on a set of users ${\cal S}$ and uses $K$-means clustering algorithm to partitions ${\cal S}$ into $K$ sets $\{{\cal S}_1, {\cal S}_2, \ldots, {\cal S}_k\}$.
%%It then selects $K$ users that from each partition that are closest to the respective cluster centers. These users are assigned the same pilot.
%%It removes the selected users from $\ncalS$ and applies the clustering and selection process again until all users are assigned the pilot.
%%
%%
%%\begin{algorithm}
%%\hrule
%%\vspace{0.05cm}
%%{\small
%% \KwInput{User locations $\Psi_u$, the set of pilots ${\cal P}$;}
%% \KwResult{Pilot assignment table ${\cal T}$;}
%% Initialization: $\Psi_u' = \Psi_u$, ${\cal T} = \emptyset$\;
%% \For{Each pilot $\nbp_k \in {\cal P}$}{
%%
%%	Obtain the $\ceil{|\Psi_u|/P}$ centroids using the $k$-means algorithm \;
%%	Find the $\Psi_{uk}$, the set of $\ceil{|\Psi_u|/P}$ in $\Psi_u'$, that are closest to each centroid \;
%%  	Assign the pilot to these users: ${\cal T} = {\cal T} \cup {\nbp_k}$\;
%%  	Remove $\Psi_{uk}$ from list of users: $\Psi_u' = \Psi_u' \setminus \nbu$\;  	
%%  	}
%% }
%% \hrulefill
%% \vspace{0.05cm}
%%\caption{The iterative $K$-means algorithm based pilot allocation.}
%%\label{algo:kmeans}
%%\end{algorithm}
%%

Both the algorithms mentioned so far do not take into account the distances among the users and the RRHs. 
As a consequence, two users that are separated by a reasonable distance, but have a common set of dominant RRHs may be assigned the same pilot.
In such a scenario, both the users will experience performance degradation. This scenario is more likely to occur when the density of RRHs is low.
In order to overcome this performance degradation, in the following section we propose a centralized RRH location aware pilot allocation algorithm.

\section{RRH Location Aware Pilot Allocation Scheme}
In this section, we present a heuristic algorithm to solve the original pilot allocation problem \eqref{eq:OptProb1} presented in Sec.~\ref{sec:RSA}. 
%Note that since the objective of the problem is to maximize sum SE, which is a function of user $\sinr$s, the pilot allocation is carried out by implicitly taking into account both RRH and user locations.
Despite a few useful constraints that we introduce to the problem to limit the size of the feasible solution space, the complexity of the problem still remains high. Hence, the practical utility of the algorithm is somewhat questionable in a large network (hundreds of users), but it provides an excellent opportunity for benchmarking any pilot allocation algorithms for a smaller network with tens of users. Further, we use this scheme to benchmark the RSA-based and max-min distance-based algorithms proposed in the previous sections.

In order to reduce the space of good quality feasible solutions, we first use a clustering algorithm to group the users that have a similar path-loss with respect to the set of RRHs. 
Once the clusters of users are obtained, we use BnP algorithm to solve the sum SE maximization problem with the additional constraint that users in the same cluster cannot be assigned the same pilot.
In the following two subsections, we discuss the clustering algorithm followed by a brief overview of the BnP algorithm with application to the problem at hand.

\subsection{RRH location-aware user clustering based on spectral graph theory}
We use a spectral graph theory-based algorithm to cluster users with similar propagation characteristics.
Before proceeding further, we present a few graph theoretic notations and definitions that are required for a rigorous exposition.

\subsubsection{Graph definitions}
Consider the weighted undirected graph $G = (\ncalV, \ncalE, \ncalW)$, where $\ncalV = \{v_1, v_2, \ldots, v_n\}$ is known as the vertex set, $\ncalE = \{e_{ij}\}_{i, j = 1, \ldots, n}$ is the edge set that contains the edges connecting these vertices, and $\ncalW = \{w_{ij}\}_{i, j = 1, \ldots, n}$ is a set of non-negative weights assigned to each edge. 
If two vertices $i, j$ are connected then $e_{ij}= 1$ and $w_{ij} > 0$. Otherwise, $e_{ij}= w_{ij} = 0$. 
The adjacency matrix of the graph $G$ is denoted by $A \in \{0, 1\}^{n \times n}$ and defined as 
$
A(i, j) = e_{ij}.
$
Further, the {\em weighted} adjacency matrix is given as $W \in \R^{n \times n}$ and defined as $W(i, j) = w_{ij}$. 
The degree matrix of a weighted graph $G$, denoted by $D \in \R^{n \times n}$, is a diagonal matrix whose $i$-th diagonal element is given as 
$
D(i, i) = \sum_{j=1}^n w_{ij}.
$
The Laplacian matrix of the graph $G$ is defined as $L = D - W$.
The $K$-cut of the graph $G$ is defined as 
\begin{align*}
\text{cut} (\ncalV_1, \ncalV_2, \ldots, \ncalV_K) = \frac{1}{2}\sum_{i=1}^K \sum_{v_l \in \ncalV_i, v_m \in \ncalV_i^C} w_{lm},
\end{align*}
where $\ncalV_i$ is the $i$-th partition of $\ncalV$.
Further, $\cup_{i=1}^K \ncalV_i = \ncalV$ and $\ncalV_i \cap \ncalV_j = \emptyset$ for $i \neq j$.
The volume of a partition $\ncalV_i$ is defined as 
\begin{align*}
\text{Vol}(\ncalV_i) = \sum_{v_j \in \ncalV_i} D(j, j).
\end{align*}
The graph $G$ is bipartite, if the vertex set can be partitioned into two sets $\ncalX, \ncalY \subset \ncalV$ such that the edges in $\ncalE$ have one end point in $\ncalX$ and another end point in $\ncalY$.
Further, the graph $G$ is a connected graph, if there is at least one path between any two vertices. 
Next, we formulate the problem of clustering users with similar propagation characteristic, namely the path-loss.

\subsubsection{Graph theoretic formulation of the clustering problem}
We consider a weighted bipartite graph where the vertices are the sets of users $\tilde{\Psi}_u = \Psi_u \cap W$ and RRHs $\tilde{\Phi}_r = \Phi_r \cap W$ over the finite spatial observation window $W$. 
An edge exists between each RRH and each user, but no edge exists among the RRHs or the users. 
The weight of an edge is the path-loss between a user and a RRH. 
In terms of the notations introduced earlier, $\ncalV = \tilde{\Psi}_u \cup \tilde{\Phi}_r$, $\ncalE = \{e_{mk} = 1 : \nbu_k \in \tilde{\Psi}_u, \nbr_m \in \tilde{\Phi}_u\}$, and $\ncalW = \{w_{mk} = \beta_{mk} : \nbu_k \in \tilde{\Psi}_u, \nbr_m \in \tilde{\Phi}_r\}$.
As mentioned earlier, $|\tilde{\Psi}_u| = N_u$ and $|\tilde{\Phi}|  = N_r$.
The degree matrix is given as 
\begin{align*}
D = \begin{bmatrix} D_u  & \boldsymbol{0}_{N_u \times N_r} \\ \boldsymbol{0}_{N_u \times N_r}^T & D_r \end{bmatrix} \in \R^{(N_u + N_r) \times (N_u + N_r)},
\end{align*}
where $D_u \in \R^{N_u \times N_u}$ is the degree matrix for the users and $D_r \in \R^{N_r \times N_r}$ is the degree matrix for the RRHs. 
Further, the weighted adjacency matrix for the considered bipartite graph is 
\begin{align*}
W = \begin{bmatrix}
\boldsymbol{0}_{N_u \times N_u} & W_{UR} \\
W_{UR}^T & \boldsymbol{0}_{N_r \times N_r}
\end{bmatrix} \in \R^{(N_u + N_r) \times (N_u + N_r)},
\end{align*}
where the rows of $W_{UR} \in \R^{N_u \times N_r}$ represent the weights associated with a user with respect to all the RRHs.
The problem of user clustering is based on the idea of partitioning the graph into desired number of groups such that edges across the groups have the lowest weights. 
In the current case, the output of the partitioning algorithm should be the clusters of users along with corresponding set of dominant RRHs such that the sum of edge weights between a set of clustered users and corresponding set of non-dominant RRHs should be minimum.
The problem can be formally stated as a {\em min-cut} problem presented below: 
\begin{align*}
& \underset{\ncalV_1, \ncalV_2, \ldots, \ncalV_K}{\text{minimize}} \quad \text{cut}(\ncalV_1, \ncalV_2, \ldots, \ncalV_K) = \underset{\ncalV_1, \ncalV_2, \ldots, \ncalV_K}{\text{minimize}} \quad \sum_{k=1}^K \frac{1}{2} \sum_{v_l \in \ncalV_k, v_m \in \ncalV_k^C} w_{lm} \\
& \text{subject to} \quad \bigcup\limits_{k=1}^K \ncalV_k =  \ncalV, \quad \bigcap\limits_{k=1}^K \ncalV_k = \emptyset,
\numberthis
\label{eq:CutOptProb}
\end{align*}
where $\ncalV_k = \tilde{\Psi}_k \cup \tilde{\Phi}_k$ contains the nodes (both users and RRHs) corresponding to the $k$-th cluster.
Different algorithms exist to solve the above min-cut problem. 
However, one major drawback of these algorithms is that they partition the vertices into unequal groups. 
To circumvent this problem, normalized ratio cut ($\mathtt{Ncut}$) is considered as the objective instead of the cut presented in~\eqref{eq:CutOptProb}~\cite{von2007tutorial}.  
Hence, the modified optimization problem can be written as 
\begin{align*}
& \underset{\ncalV_1, \ncalV_2, \ldots, \ncalV_K}{\text{minimize}} \quad \mathtt{Ncut}(\ncalV_1, \ncalV_2, \ldots, \ncalV_K) = \underset{\ncalV_1, \ncalV_2, \ldots, \ncalV_K}{\text{minimize}} \quad \sum_{k=1}^K \frac{1}{2} \frac{\sum_{v_l \in \ncalV_k, v_m \in \ncalV_k^C} w_{lm}}{\text{Vol}(\ncalV_k)}, \\
& \text{subject to} \quad \bigcup\limits_{k=1}^K \ncalV_k =  \ncalV, \quad \bigcap\limits_{k=1}^K \ncalV_k = \emptyset.
\numberthis
\label{eq:NCutOptProb}
\end{align*}
The aforementioned problem is NP-hard in nature. However, an efficient approximate solution can be obtained by relaxing the above problem that is presented next.

\subsubsection{Spectral graph theory to solve the $\mathtt{Ncut}$ problem}
The general idea of the algorithm is composed of two steps.  
In the first step, user locations are transformed into a space that captures the propagation characteristics between the set of users and the set of RRHs. 
In the second step, user clustering is performed by $K$-means algorithm to the transformed user and RRH locations.

%%\begin{figure}[!htb]
%%  \centering
%%  \includegraphics[width=0.5\columnwidth]{Figures/BiPartiteGraph}
%%  \caption{\footnotesize Illustration of a Bipartite graph.}
%%  \label{fig:BiPartiteGraph}
%%\end{figure}

For the $i$-th cluster $\ncalV_i$, let us define the vector $\nbf_i \in R^{N_t \times 1}$ whose $j$-th element is given as  
\begin{align*}
f_{ij} = 
\begin{dcases} 
 \frac{1}{\sqrt{\text{Vol}({\cal V}_i)}}, & \text{if}\  v_j \in {\cal V}_i \\
 0, & \text{if} \ v_j \in {\cal V}_i^C,
\end{dcases}
\end{align*}
where $N_t = N_u + N_r$.
Note that based on the definition of the degree matrix,  $\nbf_i^T D \nbf_i =1$.
Further, in case of the Laplacian matrix of the graph, $\nbf_i^T L \nbf_i = \text{cut}({\cal V}_i, {\cal V}_i^C)/\text{Vol}({\cal V}_i)$. 
Verifying these statements is straightforward and we refer the reader to~\cite{von2007tutorial} (and the references therein) for further insights on the graph Laplacian.

Let the matrix $F = \left[\nbf_1, \nbf_2, \ldots, \nbf_K\right]$.
Now, the optimization problem in \eqref{eq:NCutOptProb}, can be written as~
\begin{align*}
& \underset{F}{\text{minimize}} \quad \text{Tr}\left(F^TLF\right), & \text{subject to} \quad F^TDF = \nbI_K.
\numberthis
\label{eq:MatOptProb}
\end{align*}
The optimal solution for the columns of $F$ should only take discrete binary values. 
However, due to the NP-hard nature of the problem, a relaxed version of the above problem is solved, which is given as 
\begin{align*}
& \underset{F \in \R^{N_t \times K}}{\text{minimize}} \quad \text{Tr}\left(F^TLF\right), & \text{subject to} \quad F^TDF = \nbI_K.
\numberthis
\end{align*} 
Substituting $Z = D^{1/2}F$, we get 
\begin{align*}
& \underset{Z \in \R^{N_t \times K}}{\text{minimize}} \quad \text{Tr}\left(Z^TD^{-1/2}L D^{-1/2} Z\right), & \text{subject to} \quad Z^TZ = \nbI_K.
\numberthis
\label{eq:MatOptProbRlx}
\end{align*} 
Note that the above problem is convex and can be solved by reducing it to an unconstrained optimization problem using Lagrange multiplier~\cite{boyd2004convex}.
In the following lemma, the solution to the \eqref{eq:MatOptProbRlx} is presented. 
\begin{lemma}
The solution to \eqref{eq:MatOptProbRlx} consists of $K$ eigenvectors corresponding to the $K$ smallest non-zero eigenvalues of $D^{-1/2}L D^{-1/2}$.
\end{lemma}
\begin{IEEEproof}
The Lagrangian of \eqref{eq:MatOptProbRlx} is given as~\cite{boyd2004convex}
\begin{align*}
{\cal L}(Z, \Sigma) = \text{Tr}\left(Z^TD^{-1/2}L D^{-1/2} Z\right) + \text{Tr}\left(\Sigma^T (Z^T Z -\nbI_K)\right).
\end{align*}
Now, taking the derivative of ${\cal L}$ with respect to $Z$ and equating it to zero we get 
\begin{align*}
\frac{\partial {\cal L}(Z, \Sigma)}{\partial Z} = 2 D^{-1/2}L D^{-1/2} Z - 2 Z \Sigma = 0 \implies D^{-1/2}L D^{-1/2} Z = \Sigma Z. \numberthis
\label{eq:GenEigProb}
\end{align*}
Above problem is the eigenvalue problem of $D^{-1/2}L D^{-1/2}$. Let $Q$ contains the eigenvectors of $D^{-1/2}L D^{-1/2}$ and $\Sigma$ is a diagonal matrix consisting of the corresponding eigenvalues. For our solution, $Z$ contains the $K$ columns of $Q$ corresponding to the smallest $K$ eigenvalues.  
\end{IEEEproof}

Let $\tilde{Z}_n  \in \R^{N_t \times K}$ be the row normalized version of $Z$.
The transformed locations of the RRHs and users are the rows of $\tilde{Z}_n$~\cite{ng2001spectral, dhillon2001co}.
We perform K-means clustering algorithm on the rows of $\tilde{Z}_n$ to group the users and their dominant set of RRHs.
Once the cluster of users are obtained, we invoke the additional constraint of not assigning the same pilot to two users in the same cluster. 
With this additional constraint, we solve the problem~\eqref{eq:OptProb1} using BnP algorithm that is presented next. 

\subsection{Branch and price (BnP) algorithm}

\begin{figure*}[!htb]
\centering
\begin{subfigure}{0.48\textwidth}
  \centering
  \includegraphics[width=1\linewidth]{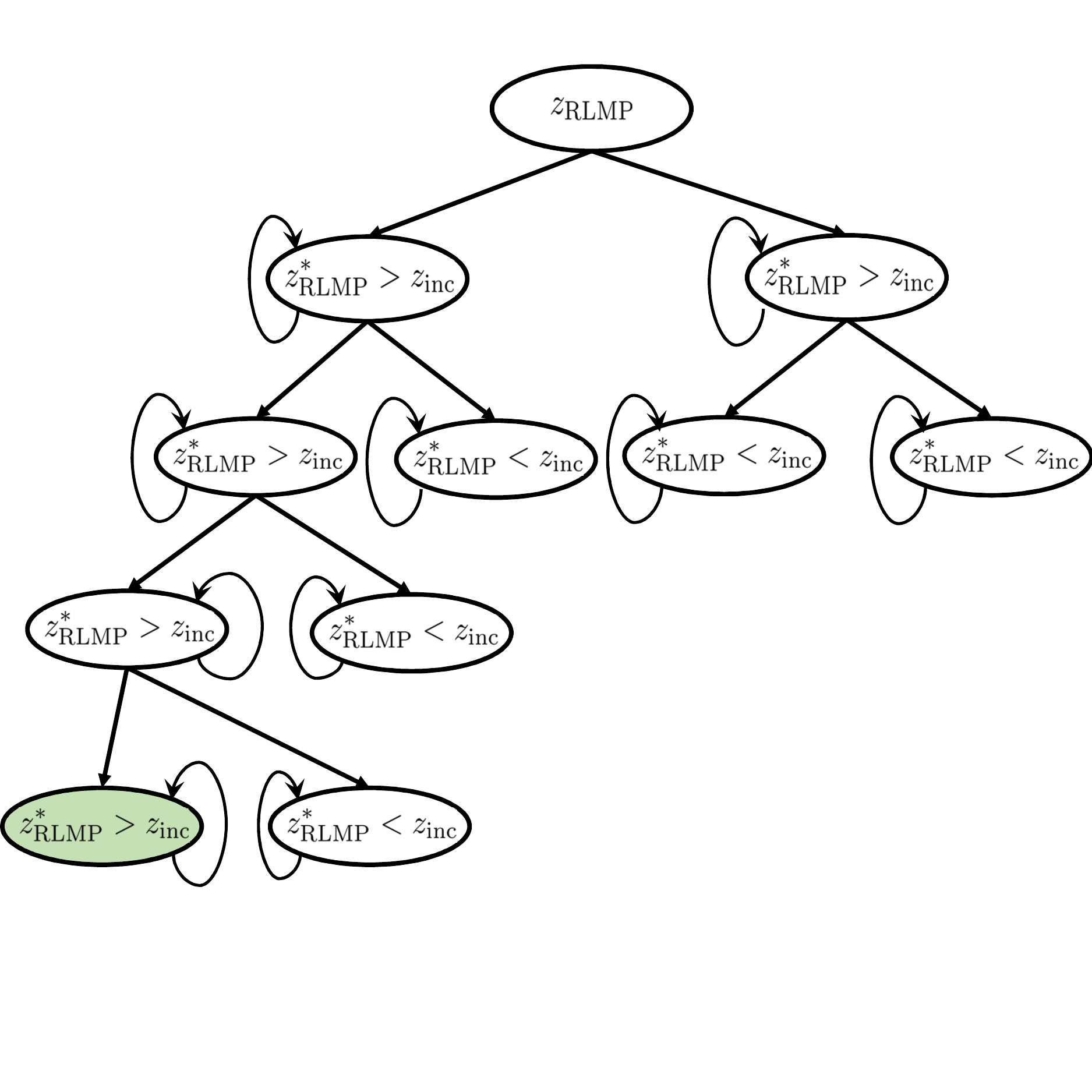}
\end{subfigure}
\begin{subfigure}{0.48\textwidth}
  \centering
  \includegraphics[width=1\linewidth]{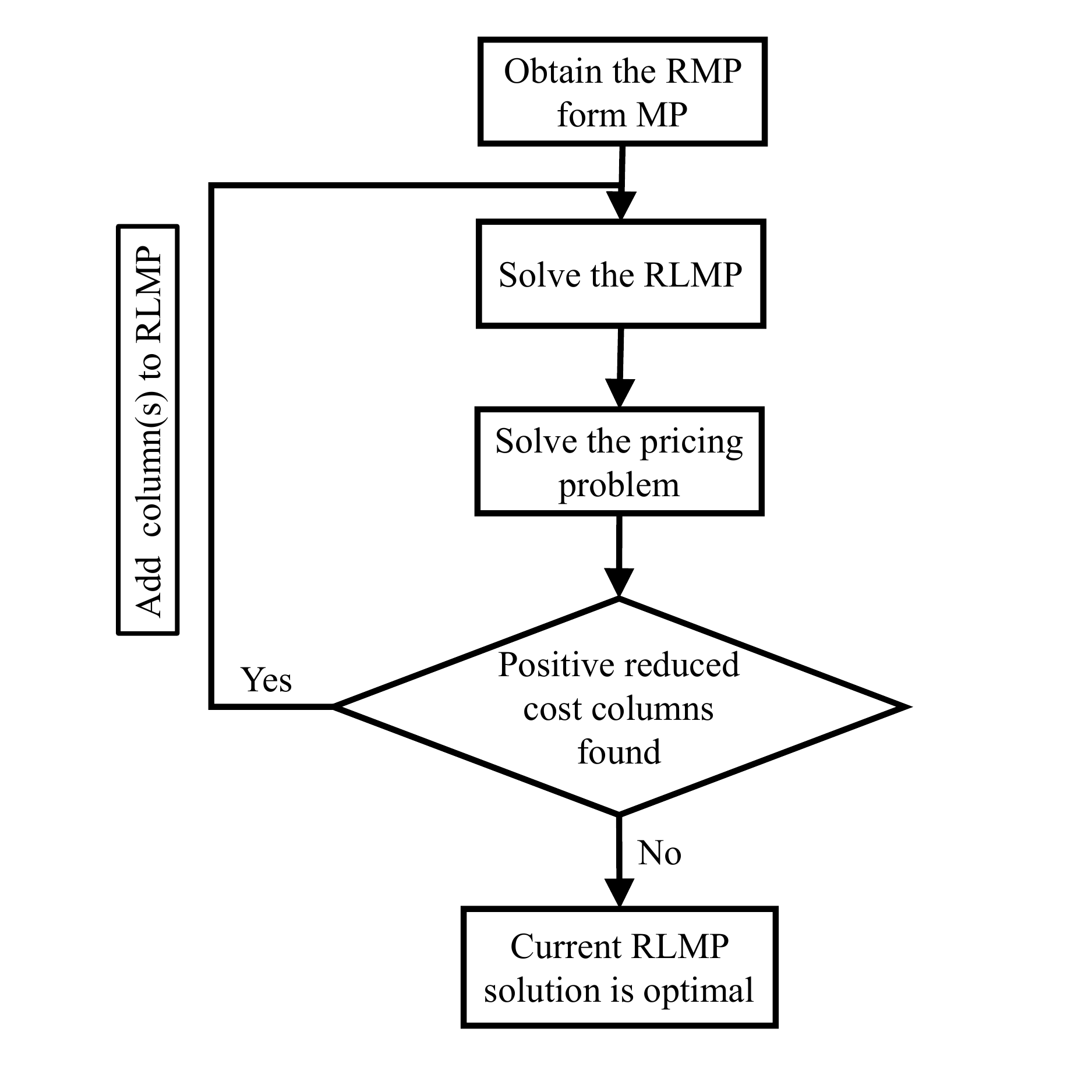}
\end{subfigure}%
\caption{\footnotesize The branch and bound tree (left). The column generation algorithm flow chart (right).}
\label{fig:BnP}
\end{figure*}

BnP is an efficient method to solve large integer programming problems and has been successfully applied to many discrete optimization problems, such as generalized assignment problem~\cite{barnhart1998branch}, graph coloring~\cite{mehrotra1996column}, and also to communication network problems of link scheduling~\cite{fu2010fast}, \cite{johansson2006cross}.
The core idea of BnP algorithm is to traverse through a branch and bound (BnB) tree. At each node of the tree, a smaller version of the original problem (by optimizing over a reduced feasible space) and a pricing problem are iteratively solved. The objective of the pricing problem is to add good quality feasible columns to the feasible space of the smaller problem. 
The process is repeated until no good quality columns are found. As a consequence of this iterative approach, a large number of (useless) columns are never considered in the entire process saving significant amount of computational resources.
Depending on the nature of the problem, some branching constraint is used to traverse through the tree.
Similar to any BnB-based algorithm, the BnP algorithm terminates once there is no improvement in the objective value in the remaining nodes of the tree compared to the incumbent solution.
An illustration of the BnP algorithm and flow of the column generation process (using the pricing problem) is presented in Fig.~\ref{fig:BnP}.

\subsubsection{Modified cost function and reduced linear master problem}
Using the clustering algorithm presented in the previous section, we get $K$ sets of clustered user partitions given by $\{\ncalV_1, \ncalV_2, \ldots, \ncalV_K\}$.
To ensure that users in the same cluster are not assigned the same pilot, we modify the cost function \eqref{eq:CostFun1} as
\begin{equation}
c(\nbx_j) = 
\begin{cases}
\sum_{\nbu_i \in \ncalA_j} \log_2\left(1 +  \Gamma_{ij}\right), & \text{if} \ \Gamma_{ij} \geq \Gamma_{\min} \ \forall \nbu_i \in \ncalA_j \\
-M, & \text{if} \ \Gamma_{ij} < \Gamma_{\min} \ \text{for any} \ \nbu_i \in \ncalA_j \\
-M, & \text{if} \ {|\ncalV_k \cap \ncalA_j| > 1 \ \text{for any}\  k = 1, 2, \ldots K},
\end{cases}
\label{eq:CostFun2}
\end{equation}
where the last row ensures that the columns that have users from the same cluster are (almost) never considered as good columns in the pricing problem.
Based on the above definition of the cost function, to make sure that the original problem remains feasible, number of users in a cluster should be less than the number of pilots. Hence, we choose $K = \max\{P, N_u/P\}$.
We call \eqref{eq:OptProb1} with the modified cost function definition as the {\em master problem (MP)}. 
Further, we refer to the problem with relaxed integer constraint \eqref{eq:IntConstr} of the MP as linear master problem (LMP), which is expressed as
\begin{subequations}
\begin{align}
\max_{\Lambda} \quad	& \sum_{s = 1}^{|\ncalA|} c(\nbx_s) \lambda_s \\ 
\text{s.t.} \quad	& A \Lambda = \boldsymbol{1} \\ 
					& \|\Lambda\|_1 = P \\ 
					& \Lambda \in [0, 1]^{|\ncalA|}.
\end{align}
\label{eq:LMP}
\end{subequations}
As mentioned earlier, at each node of the BnB tree, we solve the problem with a subset of all potential columns in $A$, and gradually keep adding good columns determined by the pricing algorithm.
We define the set $\ncalH \subset \ncalA$ and the corresponding matrix as $H$, which contains a few of the columns of $A$.
We refer to this problem as the reduced linear master problem (RLMP), which is given as 
\begin{subequations}
\begin{align}
\max_{\Lambda} \quad	& \sum_{s = 1}^{|\ncalH|} c(\nbx_s) \lambda_s \\
\text{s.t.} \quad	& H \Lambda = \boldsymbol{1} \\ \label{eq:UEpSetConstr}
					& \|\Lambda\|_1 = P \\ \label{eq:NumPilotConstr}
					& \Lambda \in [0, 1]^{|\ncalH|}.
\end{align}
\label{eq:RLMP}
\end{subequations}
Let $\Pi = [\pi_1, \pi_2, \ldots, \pi_{N_u}]$ be the set of dual variables that correspond to the constraint \eqref{eq:UEpSetConstr} and $\beta$ be the dual variable for the constraint \eqref{eq:NumPilotConstr}.
Note that the optimal set of dual variable for LMP is also the optimal set of dual variables for RLMP.

\subsubsection{Pricing problem}

At each node of the BnB tree, the RLMP is solved to optimality using any linear programming method, such as the simplex, and the corresponding dual variables are used to obtain new columns that can improve the objective of RLMP by solving a pricing problem.
The idea behind the pricing problem can be better understood from the Lagrange function of LMP, which is given as 
\begin{equation}
\ncalL(\Pi, \beta, \Lambda) = \sum_{s=1}^{|\ncalA|} c(\nbx_s) \lambda_s - \Pi^T(A \Lambda - \boldsymbol{1}) - \beta (\|\Lambda\|_1 - P).
\label{eq:LagrangeFun}
\end{equation}
Note that if a given set of solutions $\Lambda^*$ is optimal for the LMP, then the first derivative of $\ncalL$ with respect to each variable is zero. 
On the other hand, for a given set of dual variables, if we can improve the value of $\ncalL$ by increasing the value of $\lambda_j$, then it must be the case that 
\begin{equation}
\frac{\partial \ncalL(\Pi, \beta, \Lambda)}{\partial \lambda_j} = c(\nbx_j) - \Pi^T \nbx_j - \beta > 0.
\label{eq:DerivLagrange}
\end{equation}
The quantity $c(\nbx_j) - \Pi^T \nbx_j - \beta$ is known as the {\em positive reduced cost} of the column $\nbx_j$.
This provides us a direct way to add new columns to an RLMP that can improve its objective function value.
To be specific, for a given set of dual variables $(\Pi, \beta)$ corresponding to a RLMP, the pricing problem is given as 
\begin{equation}
\underset{\nbx \in A}\argmax \quad c(\nbx) - \Pi^T \nbx - \beta,
\label{eq:PricingProblem}
\end{equation}
and the optimal column is added to the RLMP, thus obtaining an augmented matrix $\tilde{H}$.
The RLMP is solved again using $\tilde{H}$ and the new set of dual variables are used in the pricing problem to get better columns. The procedure is repeated until there is no column in $A$ with positive reduced cost, i.e. $c(\nbx) - \Pi^T \nbx - \beta \leq 0$ for all the columns in $\ncalA \setminus \tilde{\ncalH}$. The flow of the column generation process is given in Fig.~\ref{fig:BnP} (right).
Note that even with the linear relaxation \eqref{eq:PricingProblem} is a non-convex non-linear problem. 
Hence, solving it to optimality in polynomial time is not possible.
However, meta-heuristic algorithms such as the genetic algorithm, or tabu search, can be used to get efficient solutions. 
In this work, we focus on solving \eqref{eq:PricingProblem} with exhaustive enumeration over the set of feasible columns. 
This process is significantly more efficient compared to solving the original problem through exhaustive enumeration.

Note that the optimal solution of the RLMP, i.e. $\Lambda^*$, is not guaranteed to be an integral solution.
The following branching rule in the BnB tree ensures that the optimal solution to the RLMP is an integer vector, thereby making $\Lambda^*$ an optimal solution to the original RMP problem that has the integrality constraint.
 
\subsubsection{Branching rule}
The objective of the branching rule is to progressively introduce branching constraints such that eventually the solution to the RLMP becomes integral~\cite{barnhart1998branch}.
The branching rule is derived from a relatively well-known result in the linear programming literature that is stated in the following lemma.
\begin{lemma}
Consider the linear maximization problem $\{\max \nbc^T \nbx: A\nbx = \boldsymbol{1}, \nbx > \boldsymbol{0}\}$. If $A$ is a totally balanced matrix, then the optimal solution $\nbx^*$ is integer valued.
\end{lemma}
For the proof along with detailed discussion of the result stated in the lemma, please refer to~\cite{hoffman2003totally}.
After solving the RLMP and pricing problem to (near) optimality, the objective is to introduce the branching constraints such that the augmented matrix of the RLMP $\tilde{H}$ eventually becomes a {\em totally balanced matrix} as we traverse through the BnB tree.  
As mentioned in~\cite{barnhart1998branch}, this can be achieved by the constraints $\tilde{h}_{pk} = \tilde{h}_{rk}$ on one branch and $\tilde{h}_{pk} = \tilde{h}_{rk} = 0$ or $\tilde{h}_{pk} \neq \tilde{h}_{rk}$ on the other branch, where $\tilde{h}_{pk}$ is the element corresponding to the $p$-th row and $k$-th column of $\tilde{H}$.
The branching constraint implicitly ensures that on one branch two users belong to the same column, while on the other branch the users belong to two different columns.
These branching constraints are introduced in the RLMP and also used in the column generation process.

\subsubsection{Node pruning and termination criterion for the BnB tree}
The backbone of the BnP algorithm is the BnB algorithm. To harvest full benefits of the BnB tree, it is essential to introduce efficient node pruning criteria so that unnecessary nodes are never visited.
Let $z_{\text{MP}}^*$, $z_{\text{LMP}}^*$ are the optimal values of the MP and LMP, respectively. 
Further, $z_{\text{LMP}}^* \leq z_{\text{RLMP}} + P c$, where $c$ is the maximum positive reduced cost of columns for a given RLMP.
When the RLMP along with pricing problem is solved to optimality, $z_{\text{LMP}}^* = z_{\text{RLMP}}^*$, since there exists no column that can improve the value of $z_{\text{RLMP}}^*$.
Let $z_{\text{inc}}$ be the incumbent solution, which is always integral in nature. 
If at a node of BnB tree, we have $z_{\text{RLMP}}^* < z_{\text{inc}}$, subsequent nodes in the branch will not provide any better solution. 
Hence, the pruning occurs at this node of the branch, i.e., subsequent nodes on the same branch are not explored and the nodes on the other branches are explored. 
Once no nodes with $z_{\text{RLMP}}^* > z_{\text{inc}}$ are found, $z_{\text{inc}}$ is the optimal solution.

\section{Results}\label{sec:Results}
In this section, through Monte Carlo simulations, we validate the theoretical analysis on pilot assignment probability and assess the performance of the RSA-inspired pilot allocation compared to other schemes presented in this work. The simulations environment for each scenario is presented in the specific subsection.

\subsection{Performance of the RSA-based pilot allocation scheme}
In this case, for the simulations, we consider a network of radius 1500 m. 
In order to avoid edge effects, points within 600 m are considered. The average user spectral efficiency is reported for the typical user located at the center.
We use the following non-line-of-sight path-loss function~\cite{series2009guidelines}:
\begin{align*}
l(d) = & 161.04 - 7.1\log_{10}(W) + 7.5\log_{10}(h) -[24.37-3.7(h/h_{\text{AP}})^2]\log_{10}(h_{\text{AP}}) \\
& +[43.42-3.1\log_{10}(h_{\text{AP}})][\log_{10}(d)-3] + 20\log_{10}(f_c)-(3.2[\log_{10}(11.75h_{\text{AT}})^2]-4.97),
\end{align*}
where $W = 20, h_{\text{AP}} = 40, h_{\text{AT}} = 1.5, h = 5, f_c = 0.45$ GHz.

\begin{figure*}[!htb]
\centering
\begin{subfigure}{0.31\textwidth}
  \centering
  \includegraphics[width=1\linewidth]{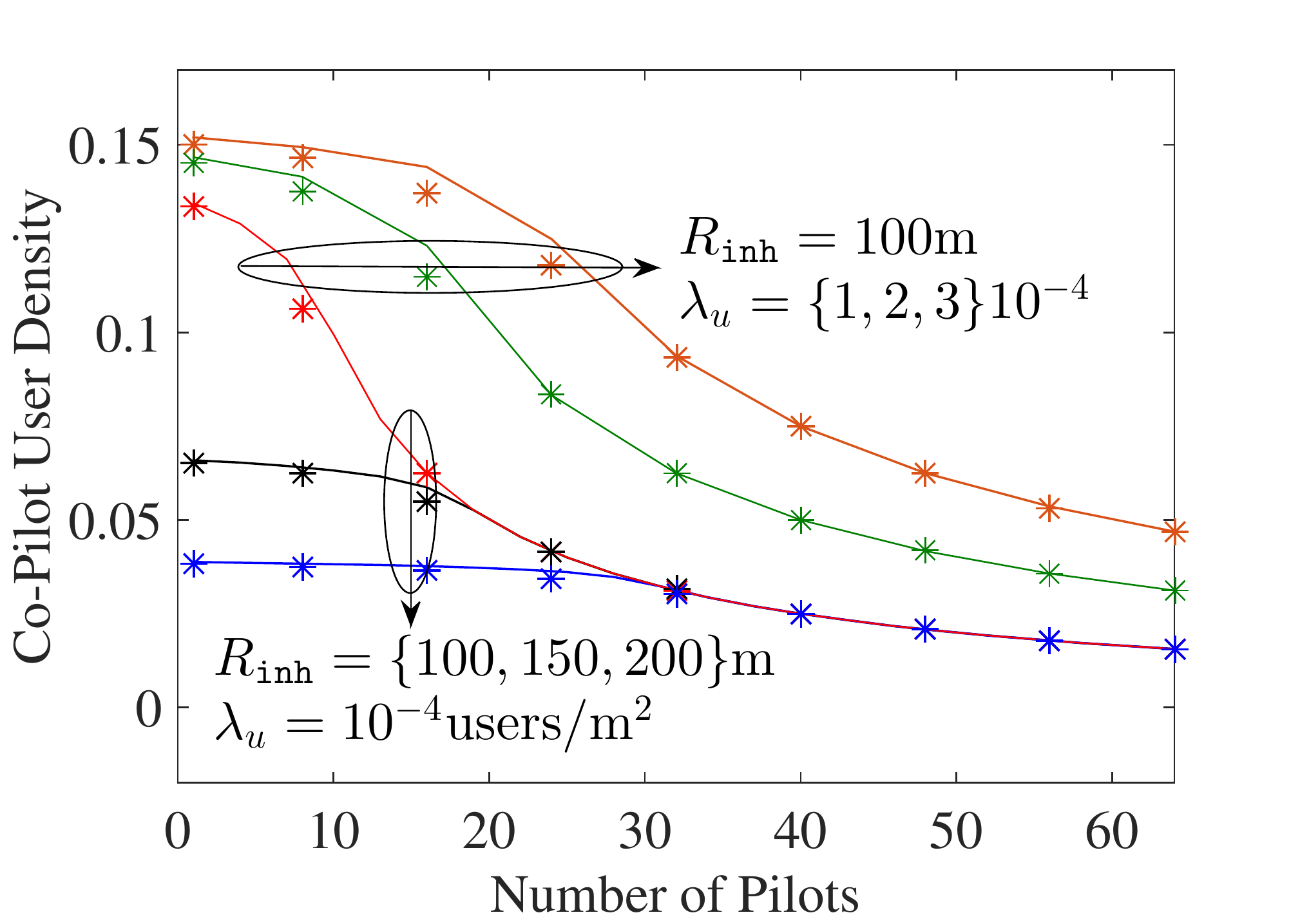}
\end{subfigure}
\begin{subfigure}{0.32\textwidth}
  \centering
  \includegraphics[width=1\linewidth]{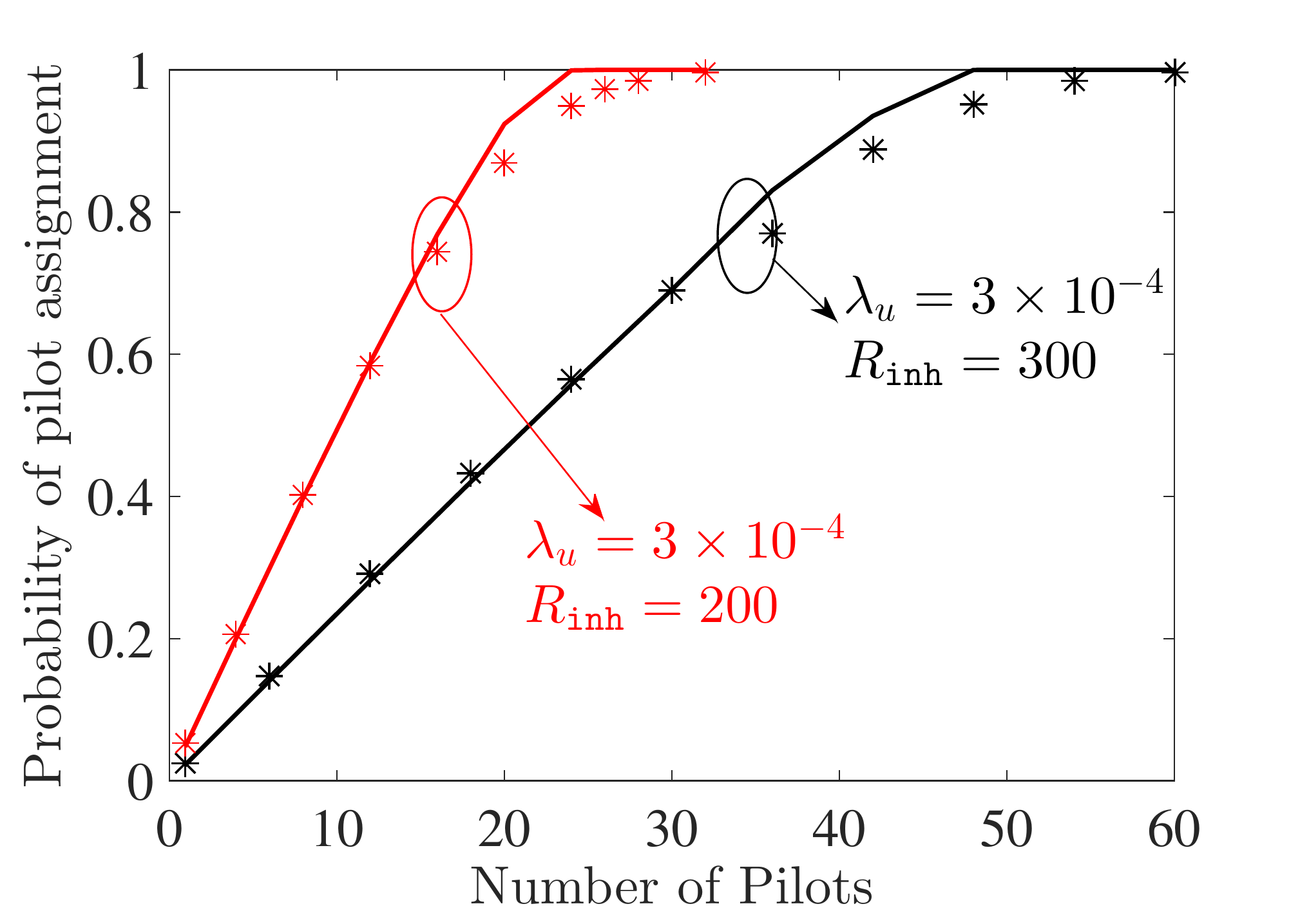}
\end{subfigure}%
\begin{subfigure}{0.32\textwidth}
  \centering
  \includegraphics[width=1\linewidth]{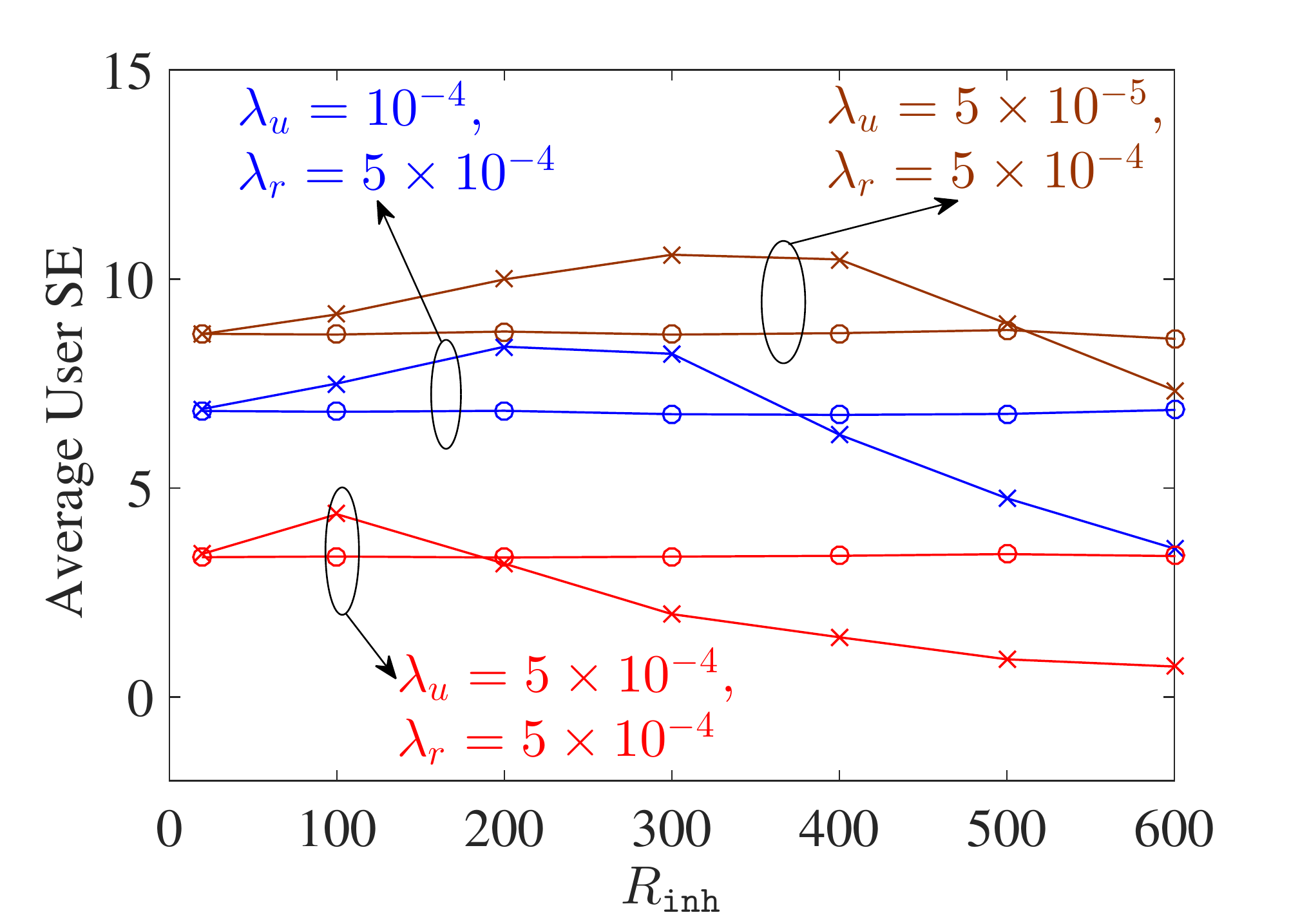}
\end{subfigure}%
\caption{\footnotesize The co-pilot user density as a function of $P$ (left), Probability of pilot assignment as a function $P$ (center), and Average user SE as a function of $R_{\tt inh}$ (right). In the first two figure, markers and solid lines represent simulations and theoretical results, respectively.}
\label{fig:RSAPerformance}
\end{figure*}

In Fig.~\ref{fig:RSAPerformance} (left), the co-pilot user density as a function of number of pilots is presented. 
As expected, the co-pilot user density decreases with increasing number of pilots.
In Fig.~\ref{fig:RSAPerformance} (center), we present the pilot assignment probability as a function of the number of pilots. 
This result is useful in determining the number of pilots that is required to achieve a certain assignment probability.
Finally, in Fig.~\ref{fig:RSAPerformance} (right), we present the average user SE as a function of $R_{\tt inh}$. To generate this result, we set the uplink pilot $\snr$ $\rho_p = 80$ dB, length of pilot sequence $\tau_p = P = 16$.
We observe that with increasing $\lambda_u$, the optimal $R_{\tt inh}$ that maximizes user SE becomes smaller.
Further, there exists a range of $R_{\tt inh}$ that provides higher user SE compared to the random pilot assignment scheme~\cite{Ngo2017}.
However, this range shrinks as $\lambda_u$ increases.

\subsection{Performance comparison of RSA-based scheme to the max-min distance-based scheme}

\begin{figure*}[!htb]
\centering
\begin{subfigure}{0.31\textwidth}
  \centering
  \includegraphics[width=1\linewidth]{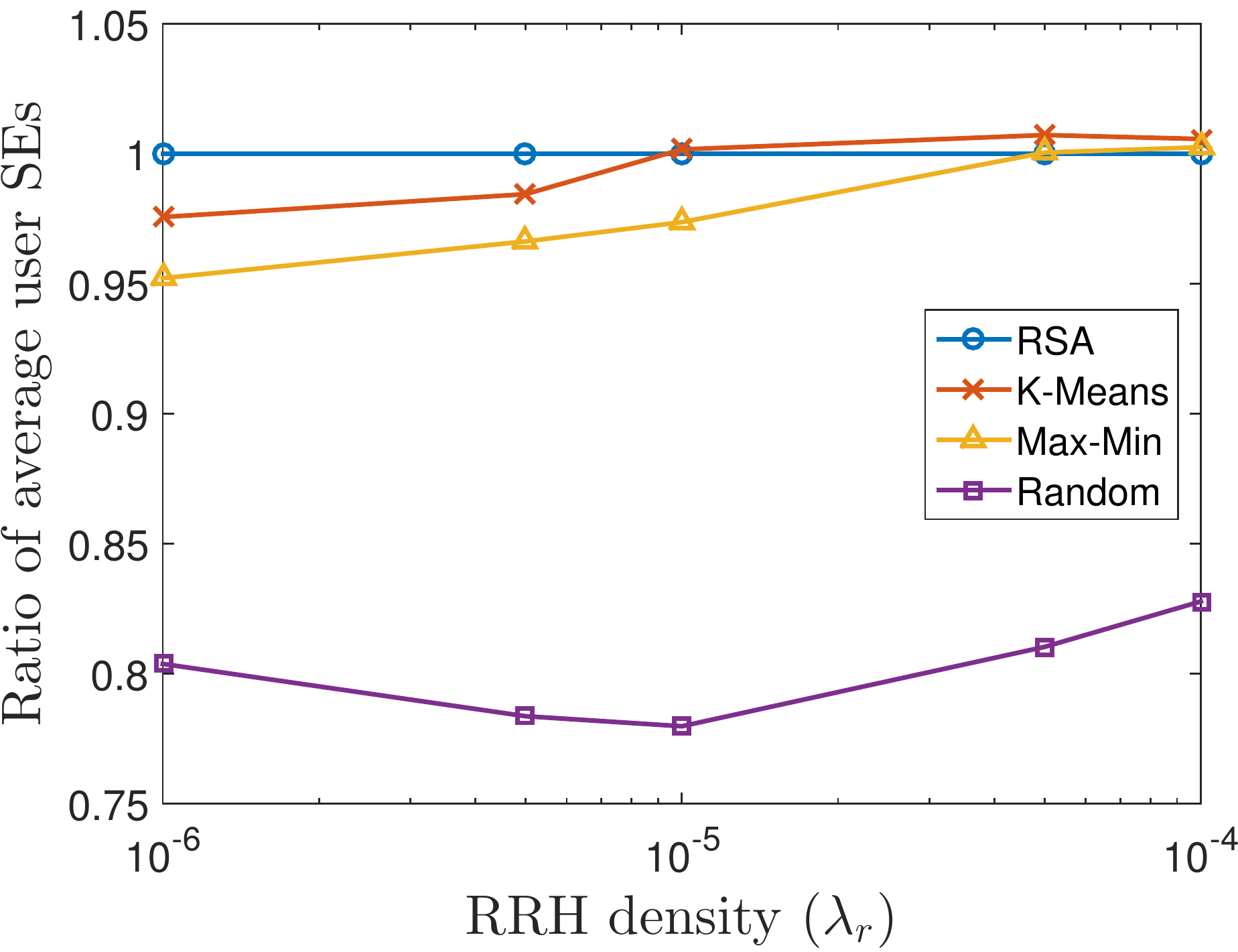}
\end{subfigure}
\begin{subfigure}{0.32\textwidth}
  \centering
  \includegraphics[width=1\linewidth]{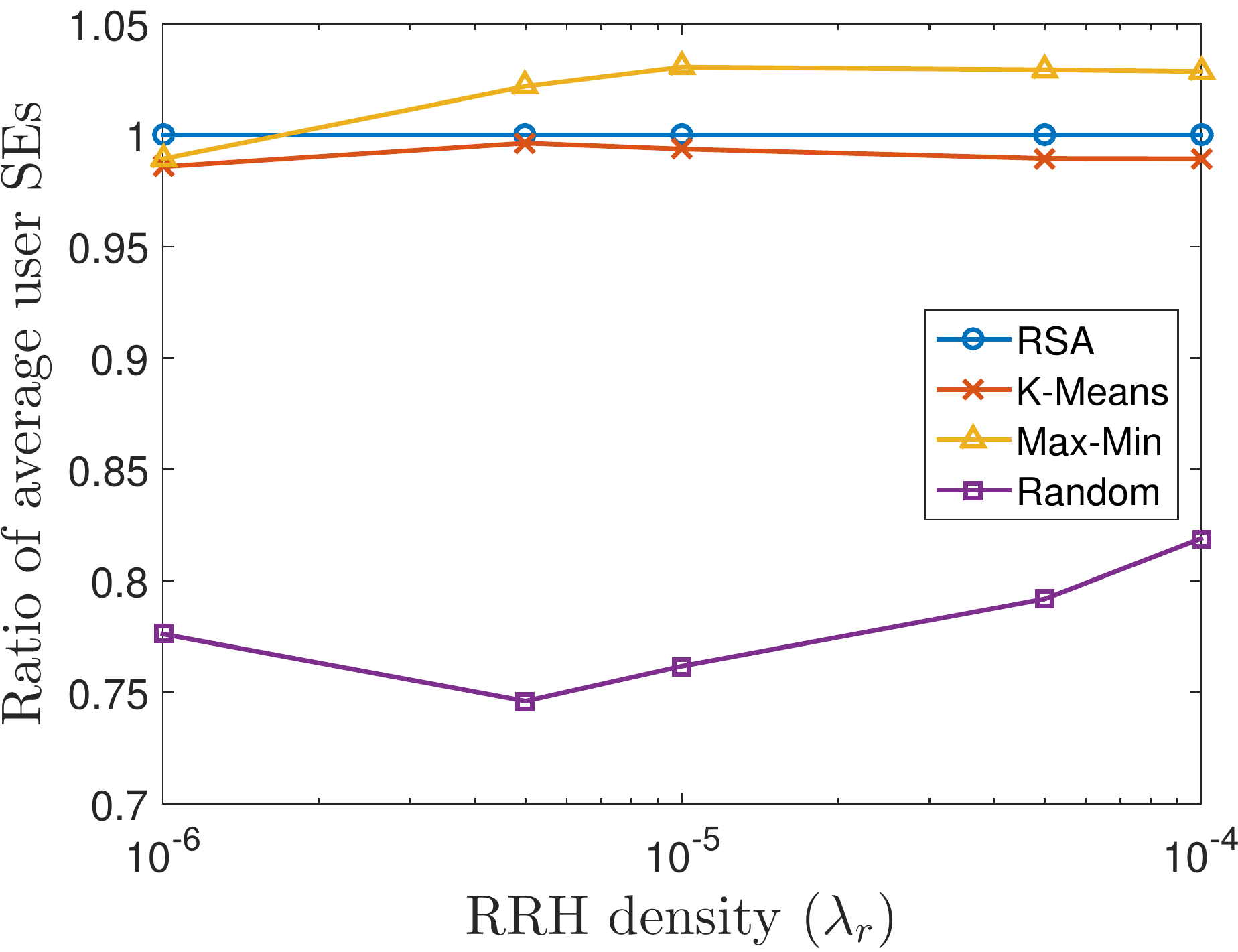}
\end{subfigure}%
\begin{subfigure}{0.32\textwidth}
  \centering
  \includegraphics[width=1\linewidth]{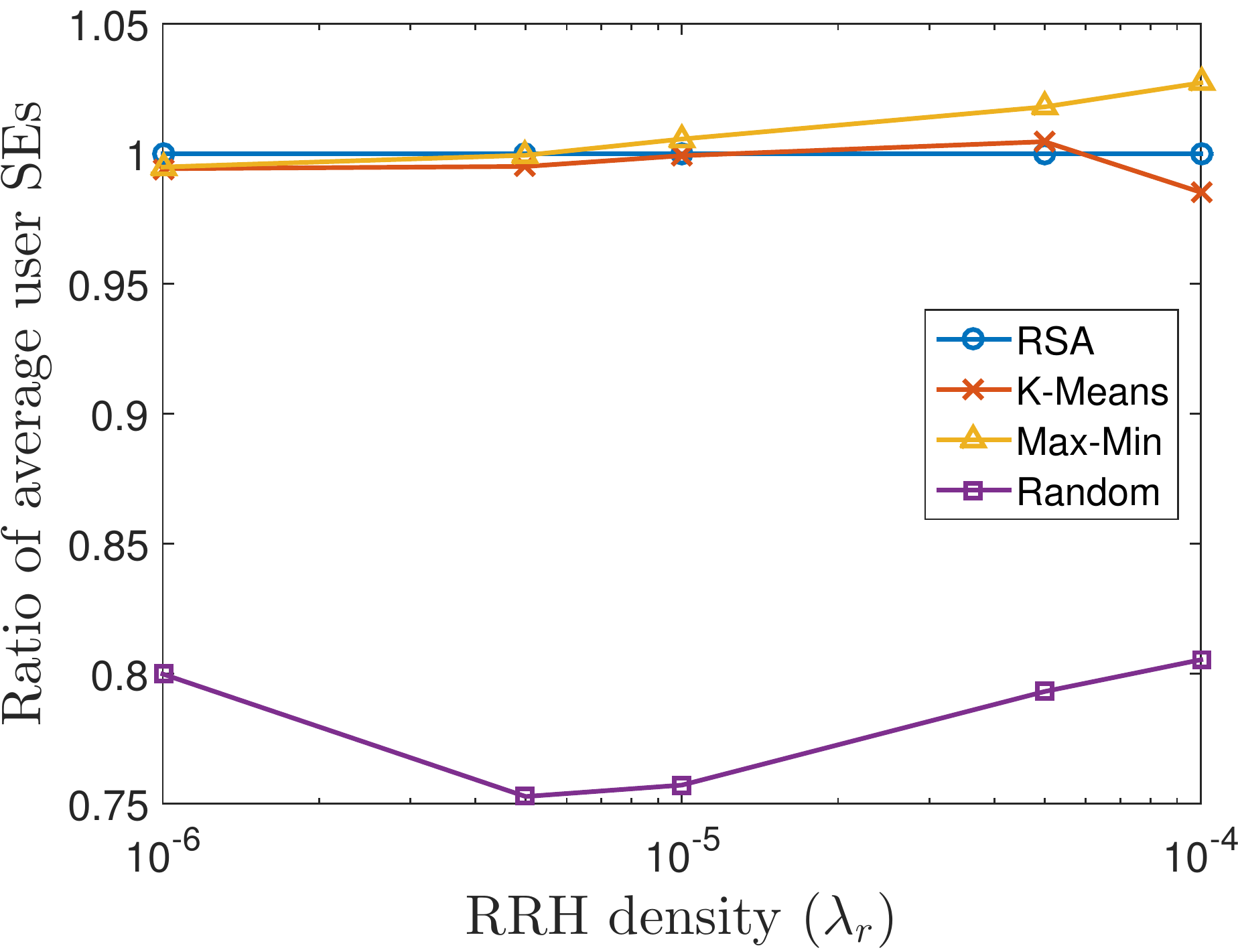}
\end{subfigure}%
\caption{\footnotesize The ratio of average user SEs of different schemes with respect to the RSA-based scheme for different system configurations: (left) $\lambda_u = 10^{-5}, P = 16$; (center)  $\lambda_u = 10^{-4}, P = 16$; (right)  $\lambda_u = 10^{-5}, P = 8$.}
\label{fig:RSAComparison}
\end{figure*}

In this subsection, we compare the performance of the RSA scheme to the max-min distance-based scheme. Further, we also provide the relative performance between RSA and the following two existing schemes in the literature: iterative K-means-based algorithm~\cite{Sabbagh2017} and the random pilot allocation algorithm~\cite{Ngo2017}. In the case of the RSA-based scheme,  for a given $\lambda_r$ and $\lambda_u$, the $R_{\tt inh}$ that maximizes the average user SE is selected.
The simulation environment remains the same as that of the previous subsection.
In Fig.~\ref{fig:RSAComparison}, we present the ratio of the average user SEs of different schemes with respect to the average user SE of the RSA scheme.
From the results, we conclude that the system performance is primarily affected by (i) the average number of users per pilot and (ii) RRH density.
When the average number of users per pilot is relatively low, the RSA scheme marginally outperforms the max-min as well as the iterative K-means algorithms, especially at the low RRH density.
On the other hand, with a relatively high average number of users per pilot, the max-min scheme performs marginally better compared to both the RSA and iterative $K$-means algorithm. In the case of the RSA, this slightly inferior performance can be attributed to the reduced pilot assignment probability in a dense environment. 
All the three schemes provide significant average user SE improvement over the random pilot allocation scheme.

\subsection{Performance comparison of the RSA-based scheme to the BnP scheme}
\begin{figure*}[!htb]
\centering
\begin{subfigure}{0.4\textwidth}
  \centering
  \includegraphics[width=1\linewidth]{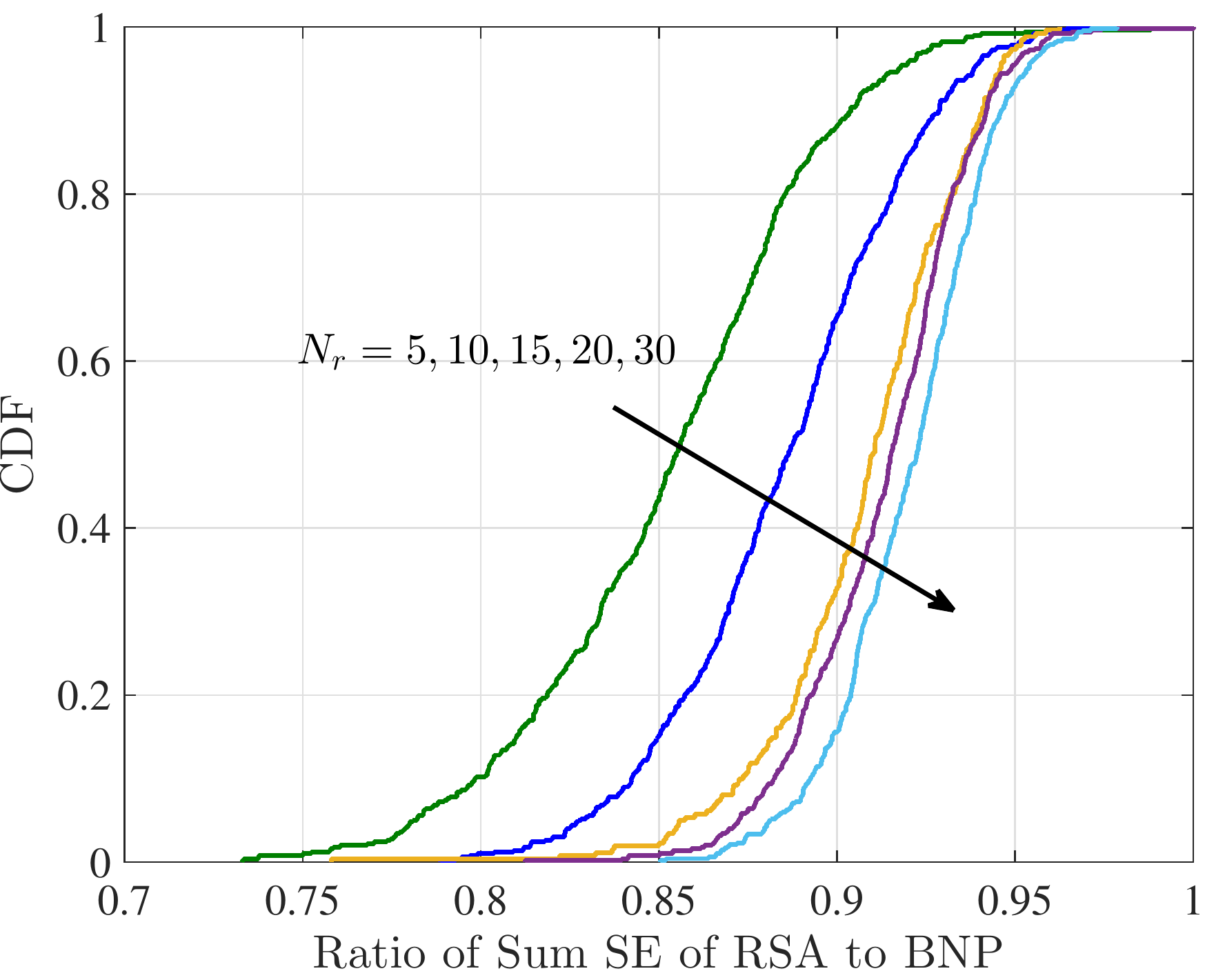}
\end{subfigure}
\begin{subfigure}{0.4\textwidth}
  \centering
  \includegraphics[width=1\linewidth]{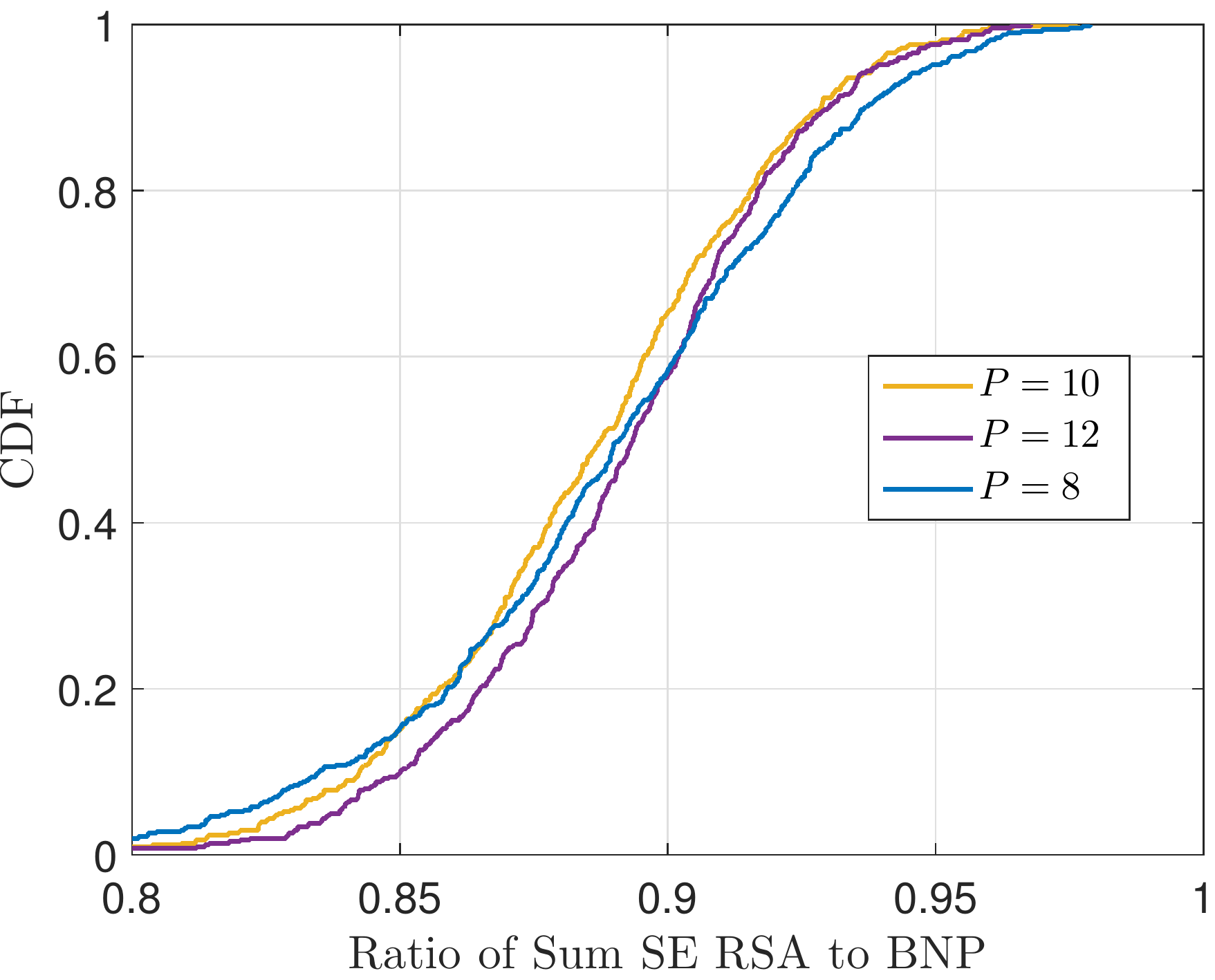}
\end{subfigure}%
\caption{\footnotesize The $\cdf$ of the ratio of sum user SE of RSA to BnP scheme for different system configuration: (left) $N_u = 48, P = 10$; (right)  $N_u = 48, N_r = 10$.}
\label{fig:RSAvsBnP}
\end{figure*}

Since the RRH location-aware heuristic scheme based on BnP algorithm exhibits significant computational complexity for a large system (hundreds of users), we compare the performance for a relatively small system with 48 users uniformly distributed over a circular area of radius 400 m. Further, these users are simultaneously served by $N_r$ RRHs distributed uniformly over the same area. The path-loss function remains the same as given in the previous subsection. 
In Fig.~\ref{fig:RSAvsBnP}, we present the cumulative distribution function ($\cdf$) of the ratio of sum user SEs for the RSA to BnP scheme.
As observed from Fig.~\ref{fig:RSAvsBnP} (left), the performance of the RSA scheme improves with the increasing number of RRHs in the system. 
However, the effect of number of pilots on the relative performance of RSA compared to the BnP scheme is negligible as evident from Fig.~\ref{fig:RSAvsBnP} (right), where RSA gives similar performance for different number of pilots.

\section{Conclusion}
In this work, we proposed a pilot assignment algorithm to mitigate the effect of pilot contamination for cell-free mMIMO systems.
Our algorithm is inspired by the RSA process, which has been used to study the adsorptions of hard particles on a surface across different scientific disciplines.
Using the well-developed analytical tools for the RSA process, we presented an accurate theoretical expression for average pilot assignment probability for the typical user in the network.
Further, the performance of the proposed algorithm was compared to two centralized pilot allocation schemes.
With respect to the first centralized scheme, which partitions the users in the network such that the minimum distance among the sets of co-pilot users is maximized, the RSA-based scheme provides competitive average user SE performance. 
The second centralized pilot allocation scheme, which is based on the BnP algorithm, provides a near-optimal solution in terms of sum user SE for a relatively small system with tens of users. 
The performance of the RSA-based scheme is appreciable with respect to the near-optimal BnP scheme.
Owing to its competitive performance and scalable distributed implementation, the RSA-based scheme is an attractive algorithm for pilot allocation in a pilot contamination limited cell-free mMIMO network.
Although technically challenging, a promising future direction of this work is to investigate a more efficient solution for the pricing problem used in the column generation process so that the BnP-based scheme can be used to benchmark the performance of even larger systems with hundreds of users.

\bibliographystyle{IEEEtran}
\bibliography{cfmMIMO_PA_Jrnl_v4}

\end{document}